\documentclass{article}
\usepackage{ijcai19}

\usepackage{times}
\usepackage{soul}
\usepackage{url}
\usepackage[hidelinks]{hyperref}
\usepackage[utf8]{inputenc}
\usepackage[small]{caption}
\usepackage{graphicx}
\usepackage{mathrsfs}
\usepackage{amssymb}
\usepackage{amsmath}
\usepackage{booktabs}
\usepackage{algorithm}
\usepackage[noend]{algorithmic}
\usepackage{comment}
\usepackage{wrapfig}
\newtheorem{mydef}{Definition}
\newtheorem{theorem}{Theorem}
\newtheorem{lemma}{Lemma}

\newtheorem{proof}{Proof}

\usepackage{empheq}
\title{Heterogeneous Gaussian Mechanism: \\ Preserving Differential Privacy in Deep Learning with Provable Robustness}

\author{
NhatHai Phan$^{1}$\footnote{Co-first authors.}\and
Minh Vu$^{5*}$ \and 
Yang Liu$^{1*}$ \and \\
Ruoming Jin$^2$\and
Dejing Dou$^{3}$\and
Xintao Wu$^4$\and 
My T. Thai$^5$\\
\affiliations
$^1$New Jersey Institute of Technology, USA;
$^2$Kent State University, USA;
$^3$University of Oregon, USA;
$^4$University of Arkansas, USA;
$^5$University of Florida, USA\\
\emails
\{phan,yl558\}@njit.edu, \{minhvu,mythai\}@ufl.edu,
rjin1@kent.edu, dou@uoregon.edu, 
xintaowu@uark.edu
}

\begin{document}

\maketitle

\begin{abstract} \vspace{-5pt}
In this paper, we propose a novel Heterogeneous Gaussian Mechanism (HGM) to preserve differential privacy in deep neural networks, with provable robustness against adversarial examples. We first relax the constraint of the privacy budget in the traditional Gaussian Mechanism from $(0, 1]$ to $(0, \infty)$, with a new bound of the noise scale to preserve differential privacy. The noise in our mechanism can be arbitrarily redistributed, offering a distinctive ability to address the trade-off between model utility and privacy loss. To derive provable robustness, our HGM is applied to inject Gaussian noise into the first hidden layer. Then, a tighter robustness bound is proposed. Theoretical analysis and thorough evaluations show that our mechanism notably improves the robustness of differentially private deep neural networks, compared with baseline approaches, under a variety of model attacks. \vspace{-5pt}
\end{abstract}


\section{Introduction}



Recent developments of machine learning (ML) significantly enhance sharing and deploying of ML models in practical applications more than ever before. This presents critical privacy and security issues, when ML models are built on personal data, e.g., clinical records, images, user profiles, etc. In fact, adversaries can conduct: 1) privacy model attacks, in which deployed ML models can be used to reveal sensitive information in the private training data \cite{Fredrikson:2015:MIA,DBLP:conf/ijcai/SW15,7958568,DBLP:journals/corr/PapernotMSW16}; and 2) adversarial example attacks \cite{DBLP:journals/corr/GoodfellowSS14} to cause the models to misclassify. Note that adversarial examples are maliciously perturbed inputs designed to mislead a model at test time \cite{DBLP:journals/corr/LiuCLS16,7958570}. That poses serious risks to deploy machine learning models in practice. Therefore, it is of paramount significance to simultaneously preserve privacy in the private training data and guarantee the robustness of the model under adversarial examples. 

To preserve privacy in the training set, recent efforts have focused on applying Gaussian Mechanism (\textbf{GM}) \cite{TCS-042} to preserve differential privacy (\textbf{DP}) in deep learning \cite{Abadi,7953387,Yu2019,Lee:2018:CDP:3219819.3220076}. The concept of DP is an elegant formulation of privacy in probabilistic terms, and provides a rigorous protection for an algorithm to avoid leaking personal information contained in its inputs. It is becoming mainstream in many research communities and has been deployed in practice in the private sector and government agencies. DP ensures that the adversary cannot infer any information with high confidence (controlled by a privacy budget $\epsilon$ and a broken probability $\delta$) about any specific tuple from the released results. 
GM is also applied to derive provable robustness against adversarial examples \cite{Lecuyer2018}. However, existing efforts only focus on either preserving DP or deriving provable robustness \cite{DBLP:journals/corr/abs-1711-00851,DBLP:journals/corr/abs-1801-09344}, but not both DP and robustness!


With the current form of GM \cite{TCS-042} applied in existing works \cite{Abadi,7953387,Lecuyer2018}, it is challenging to preserve DP in order to protect the training data, with provable robustness. In GM, random noise scaled to $\mathcal{N}(0, \sigma^2)$ is injected into each of the components of an algorithm output, where the noise scale $\sigma$ is a function of $\epsilon$, $\delta$, and the mechanism sensitivity $\Delta$. In fact, there are three major limitations in these works when applying GM: \textbf{(1)} The privacy budget $\epsilon$ in GM is restricted to $(0, 1]$, resulting in a limited search space to optimize the model utility and robustness bounds; \textbf{(2)} All the features (components) are treated the same in terms of the amount of noise injected. That may not be optimal in real-world scenarios \cite{bach-plos15,NHPhanICDM17}; and \textbf{(3)} Existing works have not been designed to defend against adversarial examples, while preserving differential privacy in order to protect the training data. These limitations do narrow the applicability of GM, DP, deep learning, and provable robustness, by affecting the model utility, flexibility, reliability, and resilience to model attacks in practice. 



\textbf{Our Contributions.} To address these issues, we first propose a novel \textit{Heterogeneous Gaussian Mechanism} (\textbf{HGM}), in which \textbf{(1)} the constraint of $\epsilon$ is extended from $(0, 1]$ to $(0, \infty)$; \textbf{(2)} a new lower bound of the noise scale $\sigma$ will be presented; and more importantly, \textbf{(3)} the magnitude of noise can be heterogeneously injected into each of the features or components. These significant extensions offer a distinctive ability to address the trade-off among model utility, privacy loss, and robustness by redistributing the noise and enlarging the search space for better defensive solutions. 

Second, we develop a novel approach, called \textbf{Secure-SGD}, to achieve both DP and robustness in the general scenario, i.e., any value of the privacy budget $\epsilon$. In Secure-SGD, our HGM is applied to inject Gaussian noise into the first hidden layer of a deep neural network. This noise is used to derive \textit{a tighter and provable robustness bound}. Then, DP stochastic gradient descent (\textbf{DPSGD}) algorithm \cite{Abadi} is applied to learn differentially private model parameters. The training process of our mechanism preserves DP in deep neural networks to protect the training data with provable robustness. To our knowledge, Secure-SGD is the first approach to learn such a secure model with a high utility. Rigorous experiments conducted on MNIST and CIFAR-10 datasets \cite{Lecun726791,krizhevsky2009learning} show that our approach significantly improves the robustness of DP deep neural networks, compared with baseline approaches. 

  \vspace{-5pt}

\section{Preliminaries and Related Work}

In this section, we revisit differential privacy, PixelDP \cite{Lecuyer2018}, and introduce our problem definition.
Let $D$ be a database that contains $n$ tuples, each of which contains data $x \in [-1, 1]^d$ and a \textit{ground-truth label} $y \in \mathbb{Z}_K$.
Let us consider a classification task with $K$ possible categorical outcomes; i.e., the data label $y$ given $x \in D$ is assigned to only one of the $K$ categories. Each $y$ can be considered as a one-hot vector of $K$ categories $y = \{y_{1}, \ldots, y_{K}\}$. 
 On input $x$ and parameters $\theta$, a model outputs class scores $f: \mathbb{R}^d \rightarrow \mathbb{R}^K$ that maps $d$-dimentional inputs $x$ to a vector of scores $f(x) = \{f_1(x), \ldots, f_K(x)\}$ s.t. $\forall k: f_k(x) \in [0, 1]$ and $\sum_{k = 1}^K f_k(x) = 1$. The class with the highest score value is selected as the \textit{predicted label} for the data tuple, denoted as $y(x) = \max_{k \in K} f_k(x)$. 
We specify a loss function $L(f(x), y)$ that represents the penalty for mismatching between the predicted values $f(x)$ and original values $y$.

\textbf{Differential Privacy.} 
The definitions of differential privacy and Gaussian Mechanism are as follows:
\begin{mydef}{$(\epsilon, \delta)$-Differential Privacy \cite{dwork2006calibrating}.} A randomized algorithm $A$ fulfills $(\epsilon, \delta)$-differential privacy, if for any two databases $D$ and $D'$ differing at most one tuple, and for all $\mathbf{o} \subseteq Range(A)$, we have:
\begin{equation} 
Pr[A(D) = \mathbf{o}] \leq e^\epsilon Pr[A(D') = \mathbf{o}] + \delta
\end{equation}
Smaller $\epsilon$ and $\delta$ enforce a stronger privacy guarantee.
\label{Different Privacy}
\end{mydef}

Here, $\epsilon$ controls the amount by which the distributions induced by $D$ and $D'$ may differ, and $\delta$ is a broken probability. DP also applies to general metrics $\rho(D, D') \leq 1$, including Hamming metric as in Definition \ref{Different Privacy} and $l_{p \in \{1, 2, \infty\}}$-norms \cite{Chatzikokolakis}. Gaussian Mechanism is applied to achieve DP given a random algorithm $A$ as follows: 
\begin{theorem}{Gaussian Mechanism \cite{TCS-042}.} Let $A: \mathbb{R}^d \rightarrow \mathbb{R}^K$ be an arbitrary $K$-dimensional function, and define its $l_2$ sensitivity to be $\Delta_{A} = \max_{D, D'} \lVert A(D) - A(D') \rVert_2$. The Gaussian Mechanism with parameter $\sigma$ adds noise scaled to $\mathcal{N}(0, \sigma^2)$ to each of the $K$ components of the output. Given $\epsilon \in (0, 1]$, the Gaussian Mechanism with $\sigma \geq \sqrt{2 \ln (1.25/\delta)}\Delta_{A}/\epsilon$ is $(\epsilon, \delta)$-DP. 
\label{Gaussian}
\end{theorem}

\textbf{Adversarial Examples.}
For some target model $f$ and inputs $(x, y_{\text{true}})$, i.e., $y_{true}$ is the true label of $x$, one of the adversary's goals is to find an \textit{adversarial example} $x^{\text{adv}} = x + \alpha$, where $\alpha$ is the perturbation introduced by the attacker, such that: \textbf{(1)} $x^{\text{adv}}$ and $x$ are close, and \textbf{(2)} the model misclassifies $x^{\text{adv}}$, i.e., $y(x^{\text{adv}}) \neq y(x)$. In this paper, we consider well-known classes of $l_{p \in \{1, 2, \infty\}}$-norm bounded attacks \cite{DBLP:journals/corr/GoodfellowSS14}. 
Let $l_p (\mu) = \{\alpha \in \mathbb{R}^d : \lVert \alpha \rVert_p \leq \mu \}$ be the $l_p$-norm ball of radius $\mu$. One of the goals in adversarial learning is to minimize the risk over adversarial examples:
\begin{equation}
\theta^* = \arg \min_{\theta} \mathbb{E}_{(x, y_{\text{true}}) \sim \mathcal{D}} \Big[\max_{\lVert \alpha \rVert_p \leq \mu} L\big(f(x + \alpha, \theta), y_{\text{true}}\big) \Big]  
\nonumber 
\end{equation}
where a specific attack is used to approximate solutions to the inner maximization problem, and the outer minimization problem corresponds to training the model $f$ with parameters $\theta$ over these adversarial examples $x^{\text{adv}}= x + \alpha$.

We revisit two basic attacks in this paper. The first one is a \textit{single-step} algorithm, in which only a single gradient computation is required. For instance, Fast Gradient Sign Method (\textbf{FGSM}) algorithm \cite{DBLP:journals/corr/GoodfellowSS14} finds an adversarial example by maximizing the loss function $L(f(x^{\text{adv}}, \theta), y_{\text{true}})$. The second one is an \textit{iterative} algorithm, in which multiple gradients are computed and updated. For instance, in \cite{DBLP:journals/corr/KurakinGB16}, FGSM is applied multiple times with small steps, each of which has a size of $\mu / T_{\mu}$, where $T_{\mu}$ is the number of steps.


\textbf{Provable Robustness and PixelDP.}
In this paper, we consider the following robustness definition. Given a benign example $x$, we focus on achieving a robustness condition to attacks of $l_p(\mu)$-norm, as follows: \vspace{-2.5pt}
\begin{equation}
\forall \alpha \in l_p(\mu): f_k(x + \alpha) > \max_{i: i\neq k} f_{i}(x + \alpha)
\label{RobustCond1} \vspace{-2.5pt}
\end{equation}
where $k$ = $y(x)$, indicating that a small perturbation $\alpha$ in the input does not change the predicted label $y(x)$.

To achieve the robustness condition in Eq. \ref{RobustCond1}, \cite{Lecuyer2018} introduce an algorithm, called \textbf{PixelDP}. By considering an input $x$ (e.g., images) as databases in DP parlance, and individual features (e.g., pixels) as tuples in DP, PixelDP shows that randomizing the scoring function $f(x)$ to enforce DP on a small number of pixels in an image guarantees robustness of predictions against adversarial examples that can change up to that number of pixels. To achieve the goal, noise $\mathcal{N}(0, \sigma^2_r)$ is injected into either input $x$ or some hidden layer of a deep neural network. That results in the following $(\epsilon_r, \delta_r)$-PixelDP condition, with a budget $\epsilon_r$ and a broken brobability $\delta_r$ of robustness, as follows:
\begin{lemma} $(\epsilon_r, \delta_r)$-PixelDP \cite{Lecuyer2018}. Given a randomized scoring function $f(x)$ satisfying $(\epsilon_r, \delta_r)$-PixelDP w.r.t. a $l_p$-norm metric, we have: 
\begin{equation}
\forall k, \forall \alpha \in l_p(\mu = 1): \mathbb{E} f_k(x) \leq e^{\epsilon_r} \mathbb{E} f_k(x + \alpha) + \delta_r
\label{PixelDPDef} 
\end{equation}
where $\mathbb{E} f_k(x)$ is the expected value of $f_k(x)$. 
\label{LemmaPixelDP}
\end{lemma}


The network is trained by applying typical optimizers, such as SGD. At the prediction time, a certified robustness check is implemented for each prediction. A generalized robustness condition is proposed as follows:\begin{equation}
\forall \alpha \in l_p(\mu = 1): \hat{\mathbb{E}}_{lb} f_k(x) > e^{2\epsilon_r} \max_{i: i\neq k} \hat{\mathbb{E}}_{ub} f_{i}(x) + (1 + e^{\epsilon_r})\delta_r
\label{RobustCon2} 
\end{equation}
where $\hat{\mathbb{E}}_{lb}$ and $\hat{\mathbb{E}}_{ub}$ are the lower bound and upper bound of the expected value $\hat{\mathbb{E}} f(x) = \frac{1}{N} \sum_N f(x)_N$, derived from the Monte Carlo estimation with an $\eta$-confidence, given $N$ is the number of invocations of $f(x)$ with independent draws in the noise $\sigma_r$.
Passing the check for a given input $x$ guarantees that \textit{no perturbation exists up to $l_p(\mu = 1)$-norm that causes the model to change its prediction result}. In other words, the classification model, based on $\hat{\mathbb{E}} f(x)$, i.e., $\arg \max_k \hat{\mathbb{E}} f_k(x)$, is consistent to attacks of $l_p(\mu = 1)$-norm on $x$ with probability $\geq \eta$. 
Group privacy \cite{dwork2006calibrating} can be applied to achieve the same robustness condition, given a particular size of perturbation $l_p(\mu)$. For a given $\sigma_r$, $\delta_r$, and sensitivity $\Delta_{p,2}$ used at prediction time, PixelDP solves for the maximum $\mu$ for which the robustness condition in Eq. \ref{RobustCon2} checks out:
\begin{empheq}[box=\fbox]{align}
& \mu_{max} = \max_{\mu \in \mathbb{R}^+} \mu \text{\ \ \ such that \ \ \ } \forall \alpha \in l_p(\mu): \nonumber
\\ 
& \hat{\mathbb{E}}_{lb} f_k(x) > e^{2\epsilon_r} \max_{i: i\neq k} \hat{\mathbb{E}}_{ub} f_{i}(x) + (1 + e^{\epsilon_r})\delta_r \nonumber 
\\
& \sigma_r = \sqrt{2 \ln (1.25/\delta_r)}\Delta_{p,2} \mu /\epsilon_r \text{\ \ and\ \ } \epsilon_r \leq 1
\label{RobustCon3}  
\end{empheq}

\section{Heterogeneous Gaussian Mechanism}

We now formally present our Heterogeneous Gaussian Mechanism (HGM) and the Secure-SGD algorithm. In Eq. \ref{RobustCon3}, it is clear that $\epsilon$ is restricted to be $(0, 1]$, following the Gaussian Mechanism (Theorem \ref{Gaussian}). That affects the robustness bound in terms of flexibility, reliability, and utility. In fact, adversaries only need to guarantee that $\hat{\mathbb{E}}_{lb} f_k(x+\alpha)$ is larger than at most $e^{2} \max_{i: i\neq k} \hat{\mathbb{E}}_{ub} f_{i}(x+\alpha) + (1 + e)\delta$, i.e., $\epsilon_r = 1$, in order to assault the robustness condition: thus, softening the robustness bound. In addition, the search space for the robustness bound $\mu_{max}$ is limited, given $\epsilon \in (0, 1]$. These issues increase the number of robustness violations, potentially degrading the utility and reliability of the robustness bound. In real-world applications, such as healthcare, autonomous driving, object recognition, etc., a flexible value of $\epsilon_r$ is needed to implement stronger and more practical robustness bounds. This is also true for many other algorithms applying Gaussian Mechanism \cite{TCS-042}.

To relax this constraint, we introduce an Extended Gaussian Mechanism as follows:
\begin{theorem}{Extended Gaussian Mechanism.} Let $A: \mathbb{R}^d \rightarrow \mathbb{R}^K$ be an arbitrary $K$-dimensional function, and define its $l_2$ sensitivity to be $\Delta_{A} = \max_{D, D'} \lVert A(D) - A(D') \rVert_2$. An Extended Gaussian Mechanism $M$ with parameter $\sigma$ adds noise scaled to $\mathcal{N}(0, \sigma^2)$ to each of the $K$ components of the output. The mechanism $M$ is $(\epsilon, \delta)$-DP, with \vspace{-5pt}
\begin{equation}
\epsilon > 0 \text{,\ \ \ } \sigma \ge \frac{\sqrt{2}\Delta_{A}}{2\epsilon} (\sqrt{s} + \sqrt{s+\epsilon}) \text{, and \ \ } s = \ln( \sqrt{ \frac{2}{\pi}} \frac{1}{\delta}) \nonumber
\end{equation} \vspace{-10pt}
\label{GGaussian}
\end{theorem}

\begin{figure}[t]
\centering
\includegraphics[width=2.1in]{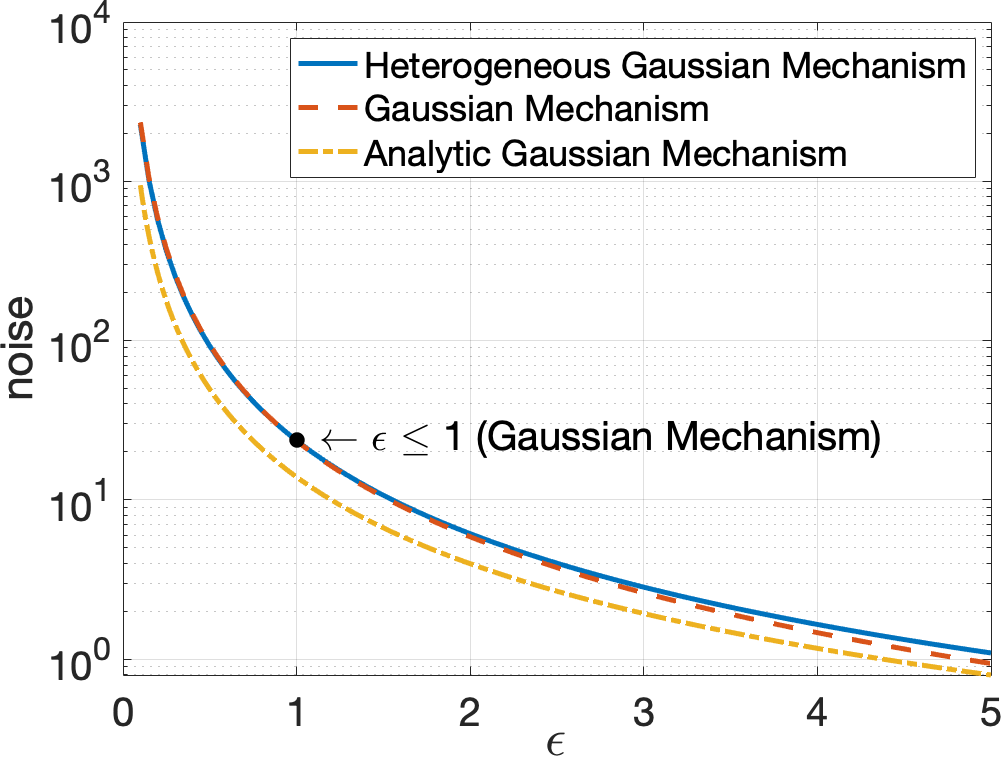} \vspace{-5pt}
\caption{The magnitude of Gaussian noise, given the traditional Gaussian Mechanism, Analytic Gaussian Mechanism, and our Heterogeneous Gaussian Mechanism.} \vspace{-5pt}
\label{proofFigure}
\end{figure}

Detailed proof of Theorem \ref{GGaussian} is in \textbf{Appendix A}\footnote{\url{https://www.dropbox.com/s/mjkq4zqqh6ifqir/HGM_Appendix.pdf?dl=0}}. The Extended Gaussian Mechanism enables us to relax the constraint of $\epsilon$. However, the noise scale $\sigma$ is used to inject Gaussian noise into each component. This may not be optimal, since different components usually have different impacts to the model outcomes \cite{bach-plos15}. To address this, we further propose a Heterogeneous Gaussian Mechanism (HGM), in which the noise scale $\sigma$ in Theorem \ref{GGaussian} can be arbitrarily redistributed. Different strategies can be applied to improve the model utility and to enrich the search space for better robustness bounds. For instance, \textit{more noise} will be injected into \textit{less important} components, or vice-versa, or even randomly redistributed. In order to achieve our goal, we introduce a noise redistribution vector $K\mathbf{r}$, where $\mathbf{r} \in \mathbb{R}^K$ that satisfies $0 \le r_i \le 1~(i \in [K])$ and $\sum_{i=1}^K r_i=1$. We show that by injecting Gaussian noise $\mathcal{N}\big(0,\sigma^2 K \mathbf{r}\big)$, where $\Delta_A = \max_{D,D'}\sqrt{\sum_{k=1}^K\frac{1}{K r_k }\big(A(D)_k-A(D')_k\big)^2}$ and $\rho(D, D') \leq 1$, we achieve $(\epsilon, \delta)$-DP. 

\begin{theorem}{Heterogeneous Gaussian Mechanism.} Let $A: \mathbb{R}^d \rightarrow \mathbb{R}^K$ be an arbitrary $K$-dimensional function, and define its $l_2$ sensitivity to be $\Delta_A = \max_{D, D'}\lVert \frac{A(D) - A(D')}{\sqrt{K\mathbf{r}}} \rVert_2 = \max_{D,D'}\sqrt{\sum_{k=1}^K\frac{1}{K r_k }\big(A(D)_k-A(D')_k\big)^2}$. A Heterogeneous Gaussian Mechanism $M$ with parameter $\sigma$ adds noise scaled to $\mathcal{N}(0, \sigma^2K\mathbf{r})$ to each of the $K$ components of the output. The mechanism $M$ is $(\epsilon, \delta)$-DP, with
\begin{equation}
\epsilon > 0 \text{,\ \ \ } \sigma \ge \frac{\sqrt{2}\Delta_{A}}{2\epsilon} (\sqrt{s} + \sqrt{s+\epsilon}) \text{, and \ \ } s = \ln( \sqrt{ \frac{2}{\pi}} \frac{1}{\delta}) \nonumber
\end{equation}
where $\mathbf{r} \in \mathbb{R}^K$ s.t. $0 \le r_i \le 1~(i \in [K])$ and $\sum_{i=1}^K r_i=1$.
\label{HGaussian}
\end{theorem}

Detailed proof of Theorem \ref{HGaussian} is in \textbf{Appendix B}$^1$. It is clear that the Extended Gaussian Mechanism is a special case of the HGM, when $\forall i \in [K]: r_i = 1/K$. Figure \ref{proofFigure} illustrates the magnitude of noise injected by the traditional Gaussian Mechanism, the state-of-the-art Analytic Gaussian Mechanism \cite{pmlr-v80-balle18a}, and our Heterogeneous Gaussian Mechanism as a function of $\epsilon$, given the global sensitivity $\Delta_A = 1$, and $\delta = 1e-5$ (a very tight broken probability), and $\forall i \in [K]: r_i = 1/K$. The lower bound of the noise scale in our HGM is just a little bit better than the traditional Gaussian Mechanism when $\epsilon \leq 1$. However, our mechanism does not have the constraint $(0, 1]$ on the privacy budget $\epsilon$. The Analytic Gaussian Mechanism \cite{pmlr-v80-balle18a}, which provides the state-of-the-art noise bound, has a better noise scale than our mechanism. However, our noise scale bound provides a distinctive ability to redistribute the noise via the vector $K\mathbf{r}$, compared with the Analytic Gaussian Mechanism. There could be numerous strategies to identify vector $\mathbf{r}$. This is significant when addressing the trade-off between model utility and privacy loss or robustness in real-world applications. In our mechanism, \textit{``more noise''} is injected into \textit{``more vulnerable''} components to improve the robustness. We will show how to compute vector $\mathbf{r}$ and identify vulnerable components in our Secure-SGD algorithm. Experimental results illustrate that, by redistributing the noise, our HGM yields better robustness, compared with existing mechanisms.

\section{Secure-SGD}

In this section, we focus on applying our HGM in a crucial and emergent application, which is enhancing the robustness of differentially private deep neural networks. Given a deep neural network $f$, DPSGD algorithm \cite{Abadi} is applied to learn $(\epsilon, \delta)$-DP parameters $\theta$. Then, by injecting Gaussian noise into the first hidden layer, we can leverage the robustness concept of PixelDP \cite{Lecuyer2018} (Eq. \ref{RobustCon3}) to derive a better robustness bound based on our HGM. 

Algorithm \ref{alg:1} outlines the key steps in our Secure-SGD algorithm. We first initiate the parameters $\theta$ and construct a deep neural network $f: \mathbb{R}^d \rightarrow \mathbb{R}^K$ (\textit{Lines 1-2}). Then, a robustness noise $\gamma \leftarrow \mathcal{N}(0, \sigma_r^2 K\mathbf{r})$ is drawn by applying our HGM (\textit{Line 3}), where $\sigma_r$ is computed following Theorem \ref{HGaussian}, $K$ is the number of hidden neurons in $h_1$, denoted as $K = |h_1|$, and $\Delta_{f}$ is the sensitivity of the algorithm, defined as the maximum change in the output (i.e., which is $h_1(x) = W^T_1 x$) that can be generated by the perturbation in the input $x$ under the noise redistribution vector $K\mathbf{r}$. 
\begin{equation}
\Delta_{f} = \max_{x, x': x \neq x'} \frac{\lVert \frac{h_1(x) - h_1(x')}{\sqrt{K\mathbf{r}}} \rVert_2}{\lVert x - x'\rVert_\infty} \leq \lVert \frac{W_1}{\sqrt{K\mathbf{r}}} \rVert_{\infty,2}
\end{equation}

\begin{algorithm}[t]
\caption{Secure-SGD}
\label{alg:1}
\begin{small}
\textbf{Input:} Database $D$, loss function $L$, parameters $\theta$, batch size $m$, learning rate $\xi_t$, gradient norm bound $C$, noise scale $\sigma$, privacy budget $\epsilon$, broken probability $\delta$, robustness parameters: $\epsilon_r$, $\delta_r$, $\Delta_f$, attack size $\mu_a$, inflation rate $\beta$, vector $\mathbf{r}$, size of the first hidden layer $K$=$|h_1|$, the number of invocations $N$
\begin{algorithmic}[1]
	\STATE \textbf{Initialize} $\boldsymbol{\theta}_0$ randomly
	\STATE \textbf{Construct} a deep neural network $f$ with \textbf{hidden layers} $\{h_1,\dots, h_O\}$, where $h_O$ is the last hidden layer
	\STATE \textbf{Draw Robustness Noise} $\gamma \leftarrow \mathcal{N}(0, \sigma_r^2 K\mathbf{r})$
	\FOR{$t \in [T]$}
        	\STATE \text{Take a random batch $B_t$ with the size $m$} \\
        	\STATE \textbf{Perturb $\forall x_i \in B_t: h_1(x_i) \leftarrow W^T_1 x_i + \gamma$}
        	\STATE \textbf{Compute Gradients}
        	\FOR{$i \in B_t$}
        		\STATE $\mathbf{g}_t(x_i) \leftarrow \nabla_{\theta_t}{L}(\boldsymbol{\theta}_t,\mathbf{x}_i)$
        	\ENDFOR
        	\STATE \textbf{Clip Gradients}
        	\FOR{$i \in B_t$}
        		\STATE $\overline{\mathbf{g}}_t(x_i) \leftarrow \mathbf{g}_t(x_i) / \max(1, \frac{\lVert \mathbf{g}_t(x_i) \rVert_2}{C})$
        	\ENDFOR
        	\STATE \textbf{Add Noise}
        	\STATE $\widetilde{g}_t \leftarrow \frac{1}{m} \big(\sum_{i}\overline{\mathbf{g}}_t(x_i) + \mathcal{N}(0, \sigma^2 C^2 \mathbf{I}) \big)$
        	\STATE \textbf{Descent}
        	\STATE $\boldsymbol{\theta}_{t+1} \leftarrow \boldsymbol{\theta}_t - \xi_t \widetilde{g}_t$
      \ENDFOR
\textbf{Output:} $(\epsilon, \delta)$-DP parameters $\boldsymbol{\theta}_T$, robust model with $(\epsilon_r, \delta_r)$ budgets
\STATE \textbf{Verified Testing:} (an input $x$, attack size $\mu_{a}$)
\STATE \textbf{Compute} robustness size $\mu_{max}$ in Eq. \ref{RobustCon4} given $x$ 
\IF{$\mu_{max} \geq \mu_a$}
	\STATE \textbf{Return} $isRobust(x) = True$, label $k$, $\mu_{max}$
\ELSE
	\STATE \textbf{Return} $isRobust(x) = False$, label $k$, $\mu_{max}$
\ENDIF
\end{algorithmic}
\end{small}
\end{algorithm}

For $l_\infty$-norm attacks, we use the following bound $\Delta_f = \sqrt{|h_1|} \lVert \frac{W_1}{K\mathbf{r}} \rVert_\infty$, where $\lVert \frac{W_1}{K\mathbf{r}} \rVert_\infty$ is the maximum 1-norm of $W_1$'s rows over the vector $K\mathbf{r}$. The vector $\mathbf{r}$ can be computed as the forward derivative of $h_1(x)$ as follows:
\begin{equation}
\mathbf{r} = \frac{\mathbf{s}}{\sum_{s_i \in \mathbf{s}} s_i}, \textit{\ where\ } \mathbf{s} = \frac{1}{n} \sum_{x \in D} \Big|\frac{\partial L(\theta, x)}{\partial h_1(x)} \Big|^\beta
\end{equation}
where $\beta$ is a user-predefined inflation rate.
It is clear that features, which have higher forward derivative values, will be more vulnerable to attacks by maximizing the loss function $L(\theta, x)$. These features are assigned larger values in vector $\mathbf{r}$, resulting in more noise injected, and vice-versa. The computation of $\mathbf{r}$ can be considered as a prepossessing step using a pre-trained model. It is important to note that the utilizing of $\mathbf{r}$ does not risk any privacy leakage, since $\mathbf{r}$ is only applied to derive provable robustness. It does not have any effect on the DP-preserving procedure in our algorithm, as follows. First, at each training step $t \in T$, our mechanism takes a random sample $B_t$ from the data $D$, with sampling probability $m/n$, where $m$ is a batch size (\textit{Line 5}). For each tuple ${x}_i \in B_t$, the first hidden layer is perturbed by adding Gaussian noise derived from our HGM (\textit{Line 6, Alg. \ref{alg:1}}):
\begin{equation}
h_1(x_i) = W^T_1 x_i + \gamma
\end{equation}

This ensures that the scoring function $f(x)$ satisfies $(\epsilon_r, \delta_r)$-PixelDP (Lemma \ref{PixelDPDef}). Then, the gradient $\mathbf{g}_t({x}_i) = \nabla_{\theta_t}{L}(\boldsymbol{\theta}_t,{x}_i)$ is computed (\textit{Lines 7-9}). The gradients will be bounded by clipping each gradient in $l_2$ norm; i.e., the gradient vector $\mathbf{g}_t(x_i)$ is replaced by $\mathbf{g}_t(x_i)/\max(1, \lVert \mathbf{g}_t(x_i) \rVert_2/C)$ for a predefined threshold $C$ (\textit{Lines 10-12}). Uniformed normal distribution noise is added into gradients of parameters $\boldsymbol{\theta}$ (\textit{Line 14}), as:
\begin{equation}
\widetilde{g}_t \leftarrow \frac{1}{m} \Big(\sum_{i}\frac{\mathbf{g}_t({x}_i)}{\max(1, \frac{\lVert \mathbf{g}_t({x}_i)^2 \rVert}{C})} + \mathcal{N}(0, \sigma^2 C^2 \mathbf{I}) \Big)
\end{equation}

\begin{figure*}[t]
\centering
$\begin{array}{c@{\hspace{0.0in}}c@{\hspace{0.0in}}c@{\hspace{0.0in}}c}
\includegraphics[width=1.75in]{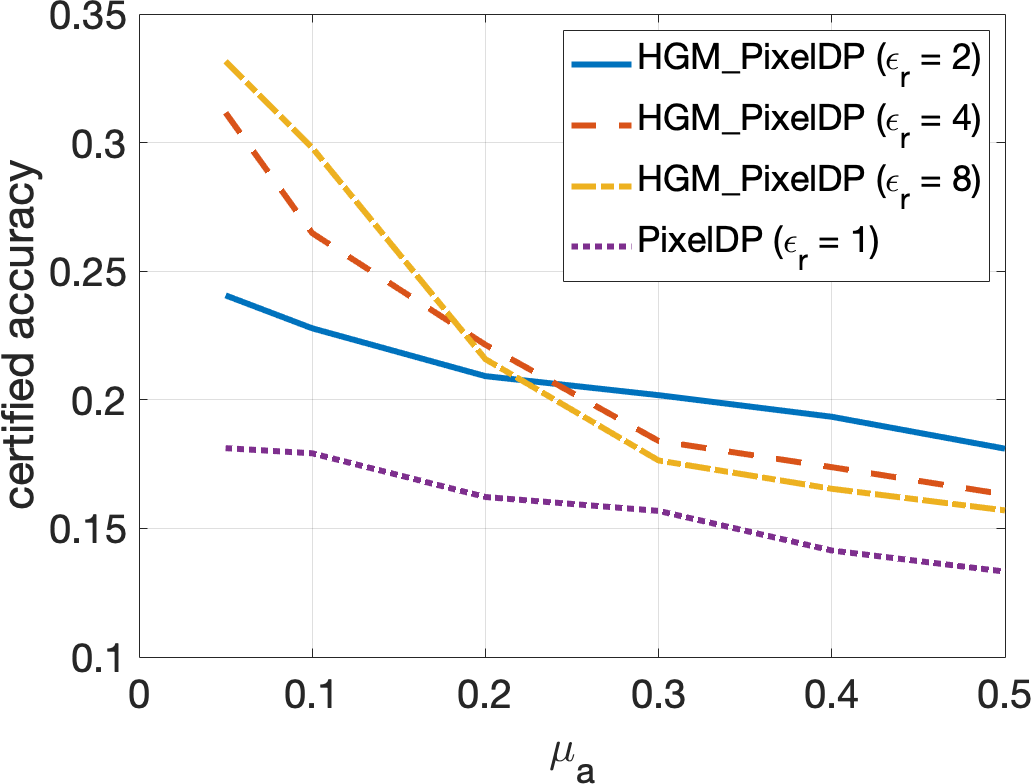} & \includegraphics[width=1.75in]{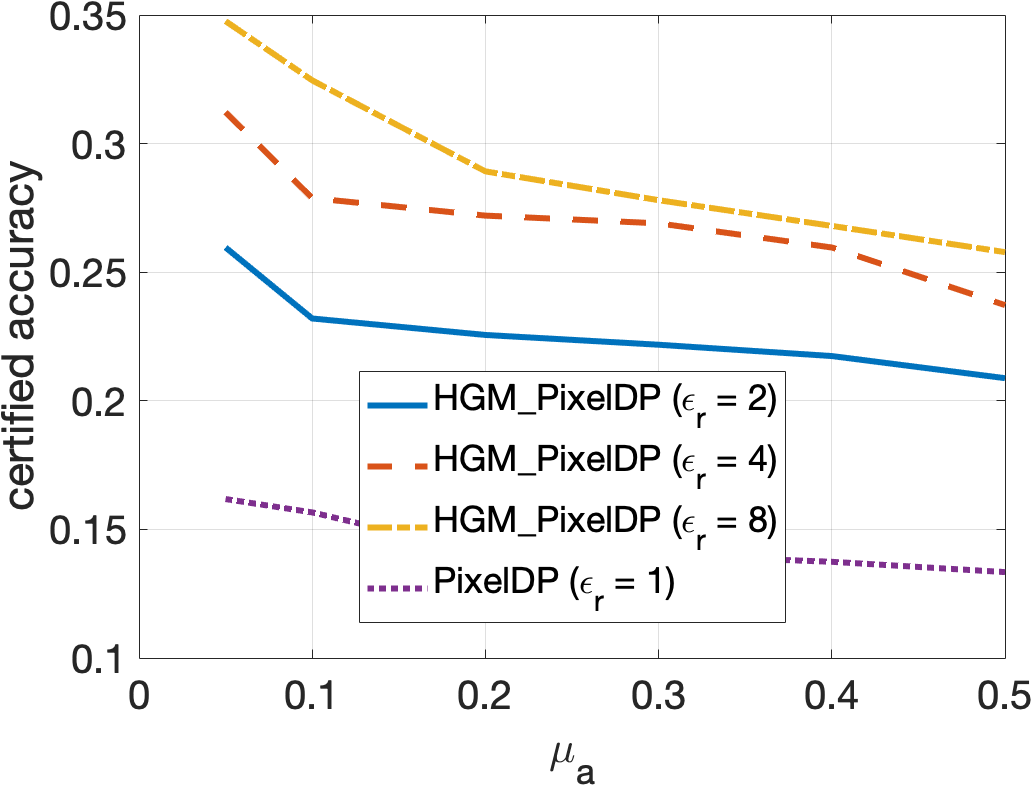} & \includegraphics[width=1.75in]{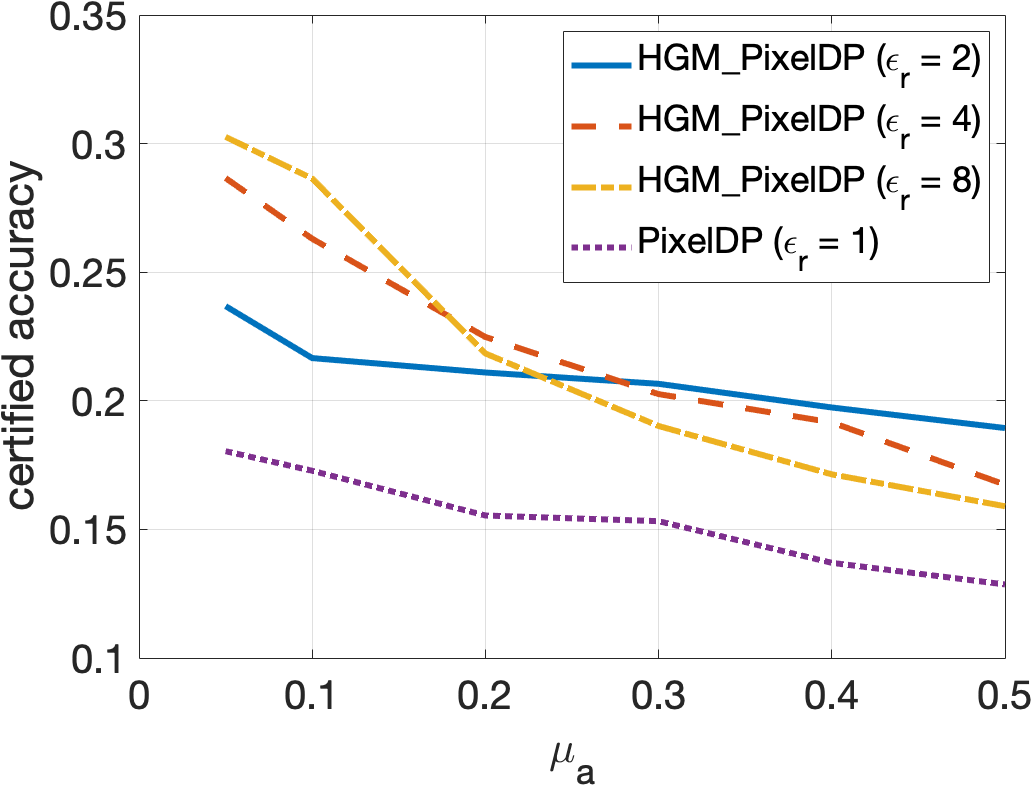} & \includegraphics[width=1.75in]{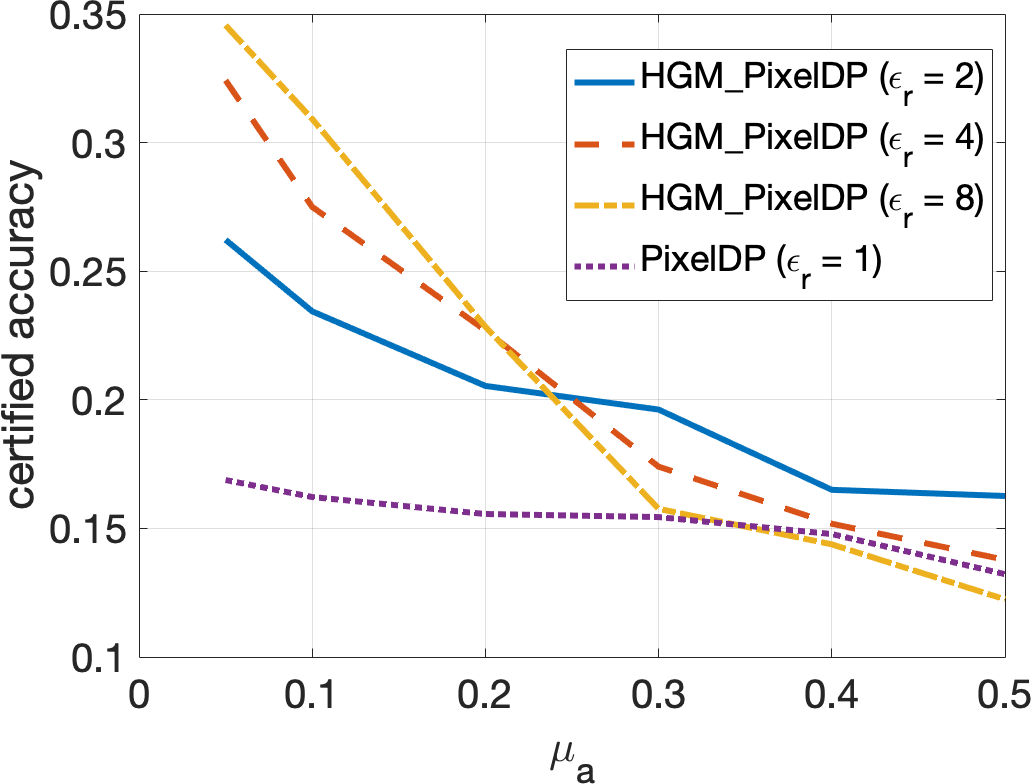} \\ [0.0cm]
\mbox{(a) I-FGSM attacks} & \mbox{(b) FGSM attacks} & \mbox{(c) MIM attacks} & \mbox{(d) MadryEtAl attacks}
\end{array}$ \vspace{-7.5pt}
\caption{Certified accuracy on the CIFAR-10 dataset, given HGM\_PixelDP and PixelDP (i.e., \textbf{no DP preservation}).} \vspace{-8.5pt}
\label{Cifar1}
\end{figure*}

\begin{figure*}[t]
\centering
$\begin{array}{c@{\hspace{0.0in}}c@{\hspace{0.0in}}c@{\hspace{0.0in}}c}
\includegraphics[width=1.75in]{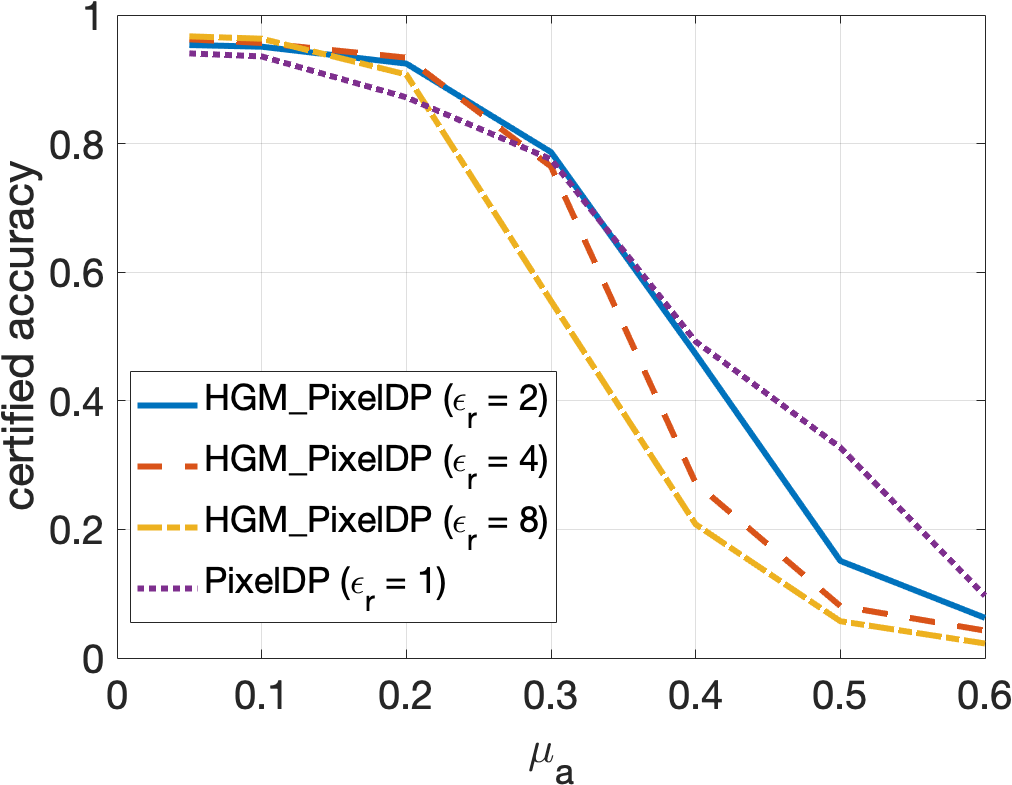} & \includegraphics[width=1.77in]{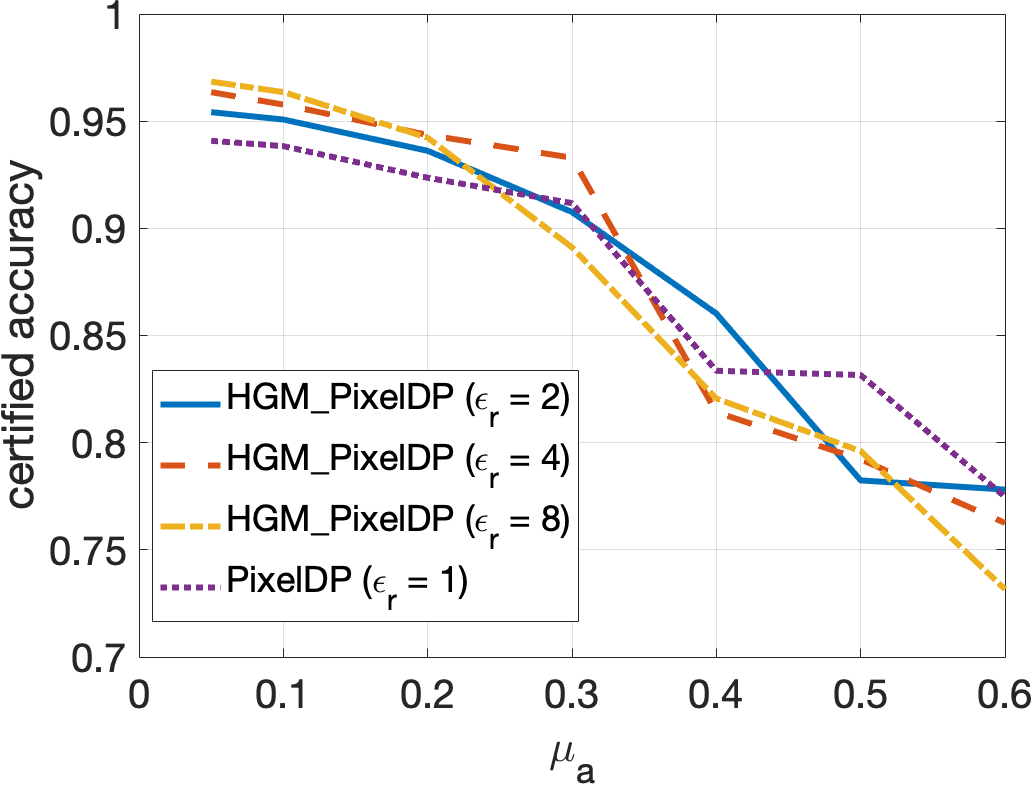} & \includegraphics[width=1.75in]{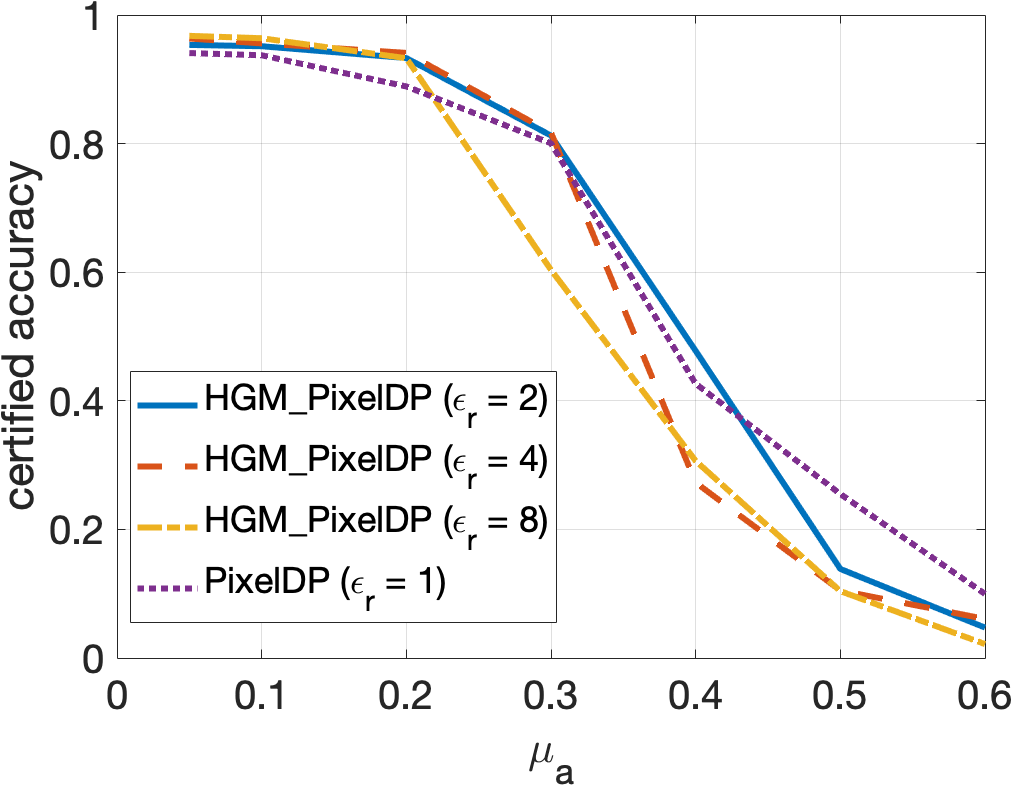} & \includegraphics[width=1.75in]{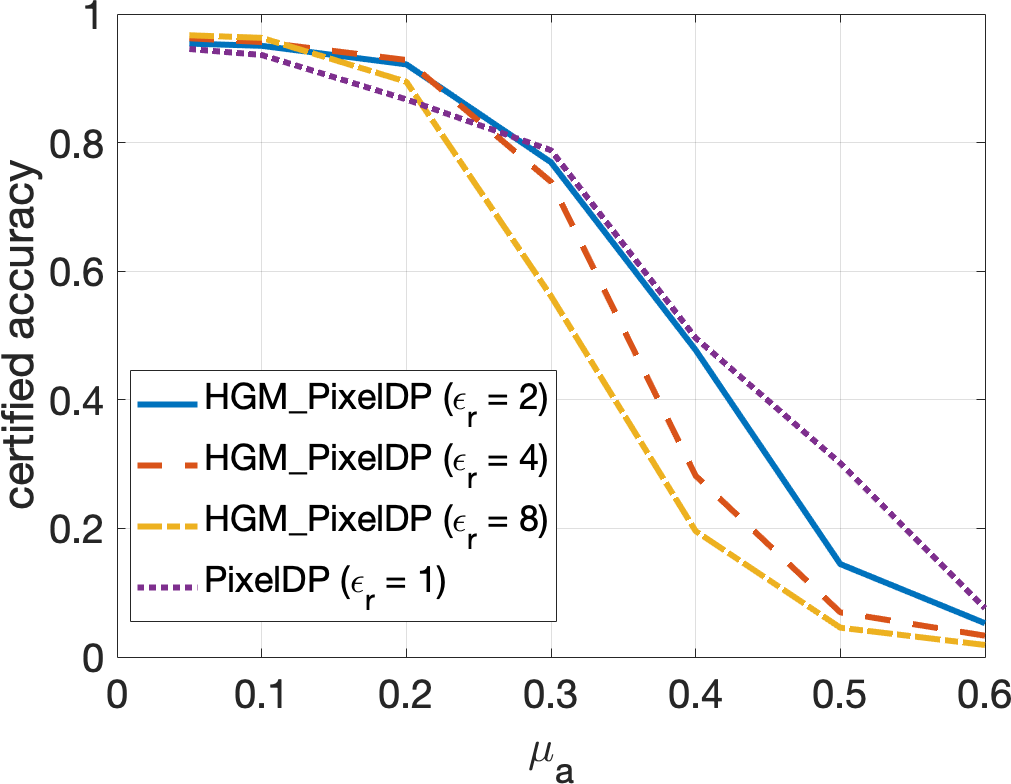} \\ [0.0cm]
\mbox{(a) I-FGSM attacks} & \mbox{(b) FGSM attacks} & \mbox{(c) MIM attacks} & \mbox{(d) MadryEtAl attacks}
\end{array}$ \vspace{-7.5pt}
\caption{Certified accuracy on the MNIST dataset, given HGM\_PixelDP and PixelDP (i.e., \textbf{no DP preservation}).} \vspace{-8.5pt}
\label{MNIST1}
\end{figure*}

\begin{figure*} [!t]
\centering
$\begin{array}{c@{\hspace{0.0in}}c@{\hspace{0.0in}}c@{\hspace{0.0in}}c}
\includegraphics[width=1.77in]{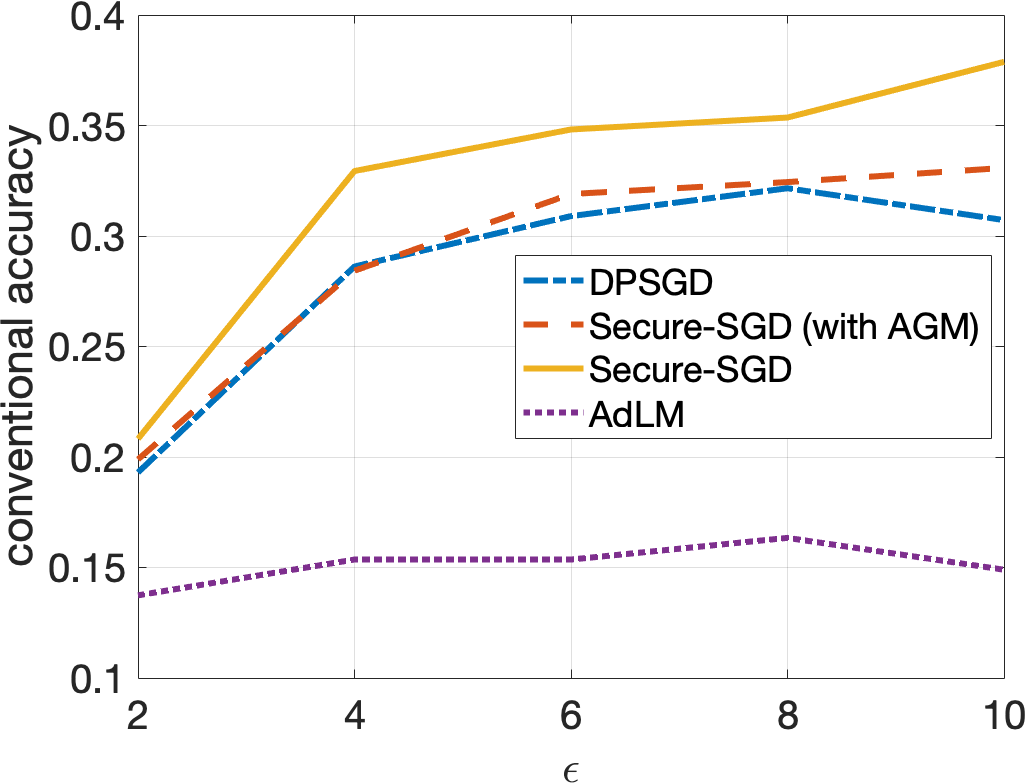} & \includegraphics[width=1.75in]{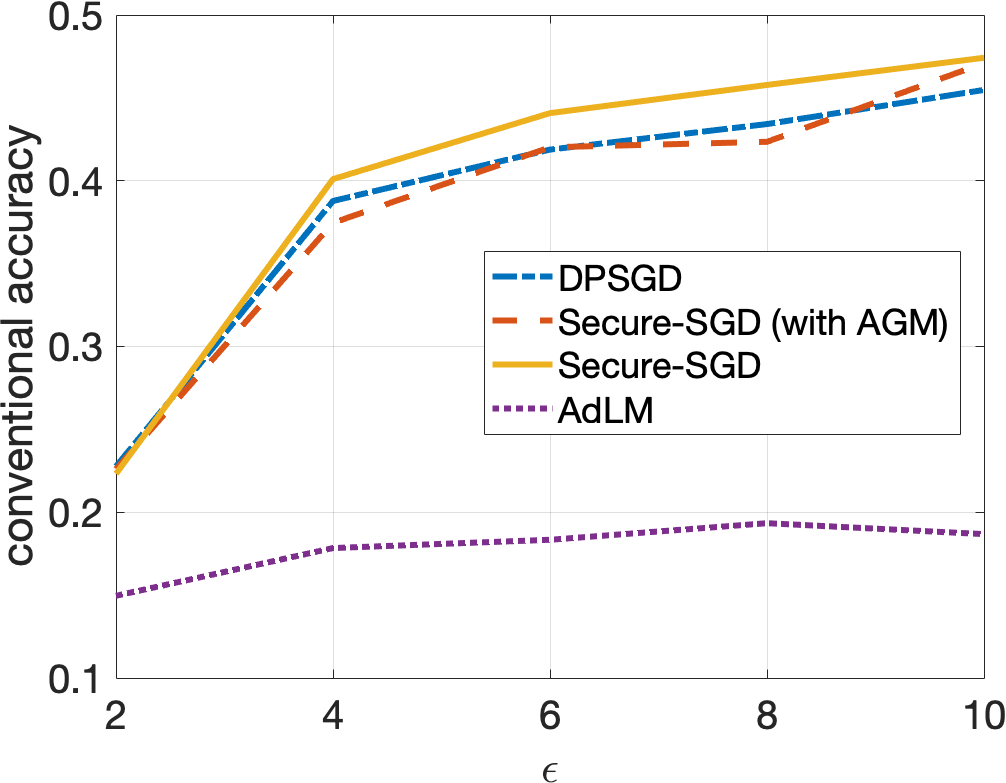} & \includegraphics[width=1.76in]{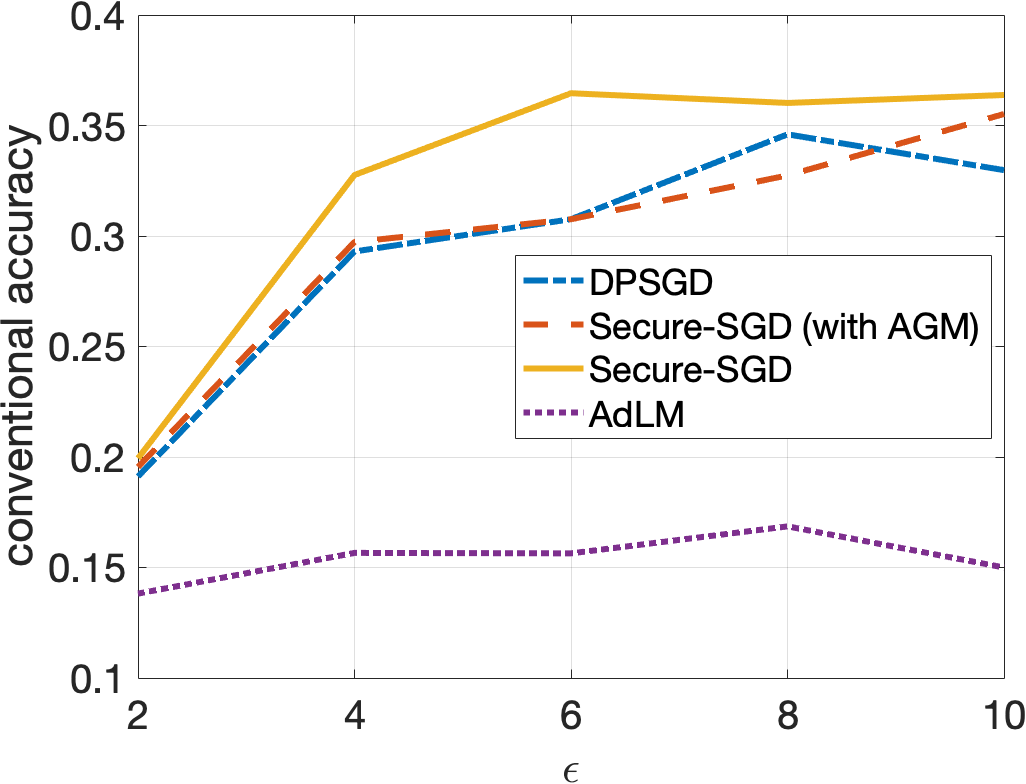} & \includegraphics[width=1.76in]{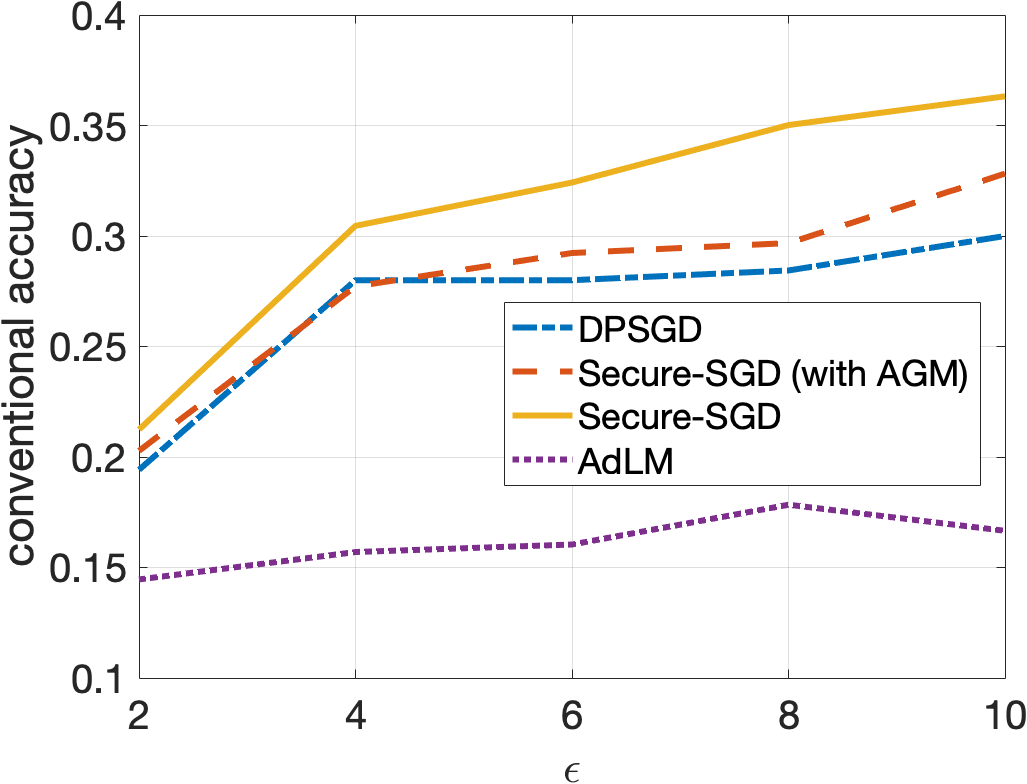} \\ [0.0cm]
\mbox{(a) I-FGSM attacks} & \mbox{(b) FGSM attacks} & \mbox{(c) MIM attacks} & \mbox{(d) MadryEtAl attacks}
\end{array}$ \vspace{-7.5pt}
\caption{Conventional accuracy on the CIFAR-10 dataset, given Secure-SGD, DPSGD, and AdLM, i.e., $l_\infty(\mu_a = 0.2)$, $\epsilon_r = 8$.} \vspace{-8.5pt}
\label{Cifar2}
\end{figure*}

\begin{figure*}[!t]
\centering
$\begin{array}{c@{\hspace{0.0in}}c@{\hspace{0.0in}}c@{\hspace{0.0in}}c}
\includegraphics[width=1.75in]{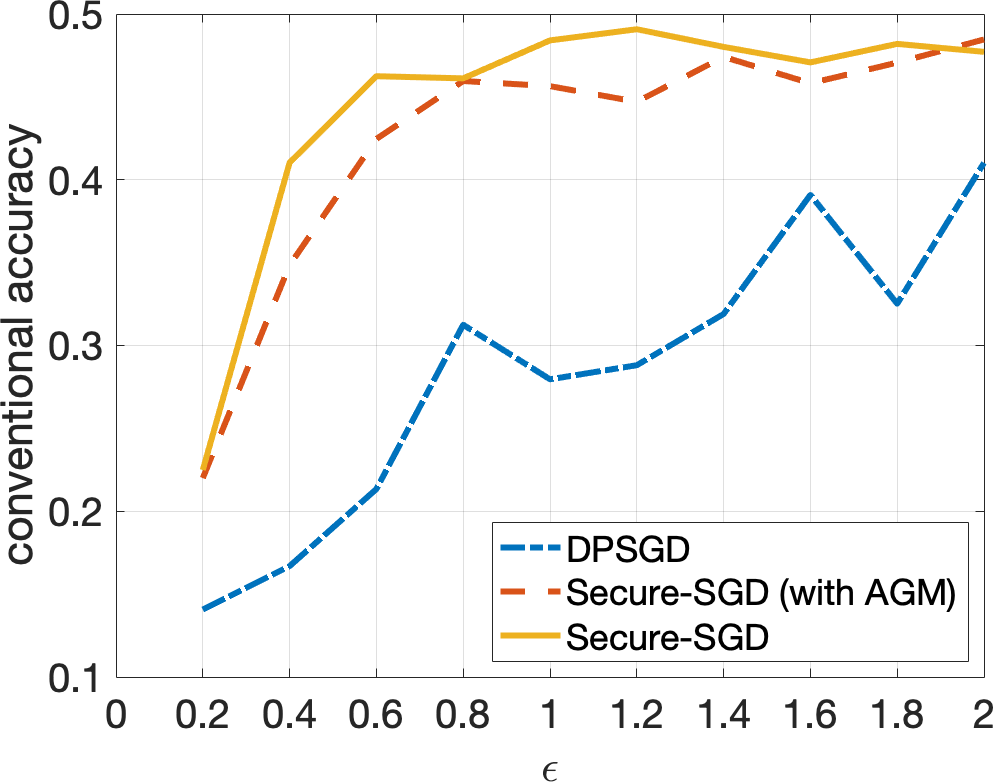} & \includegraphics[width=1.75in]{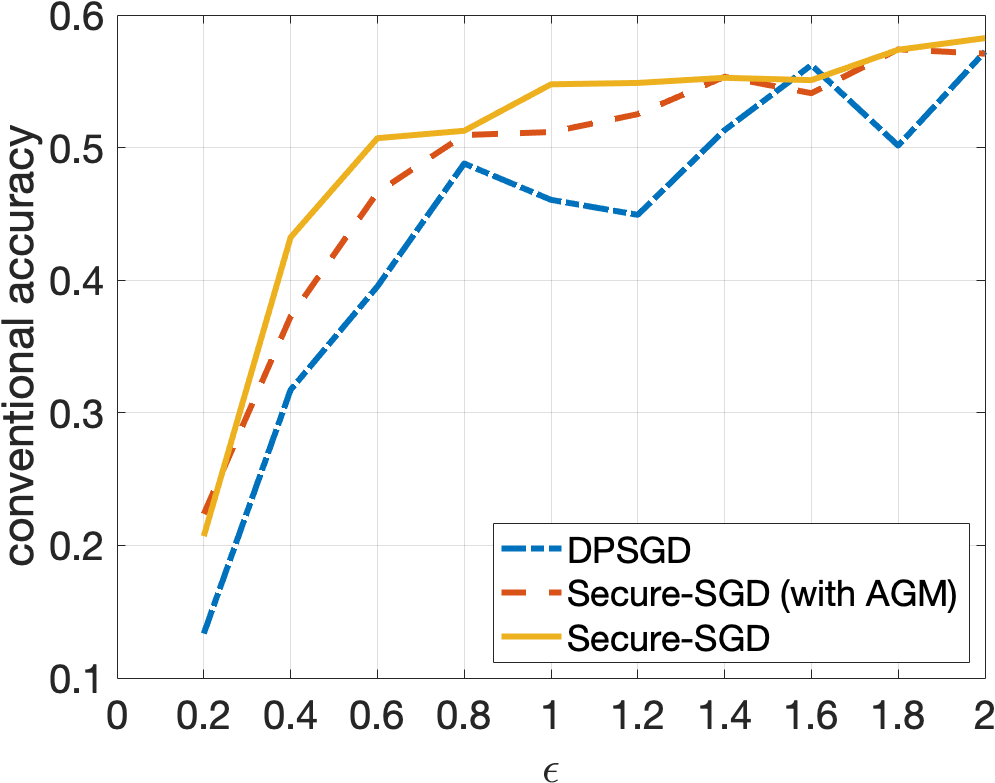} & \includegraphics[width=1.75in]{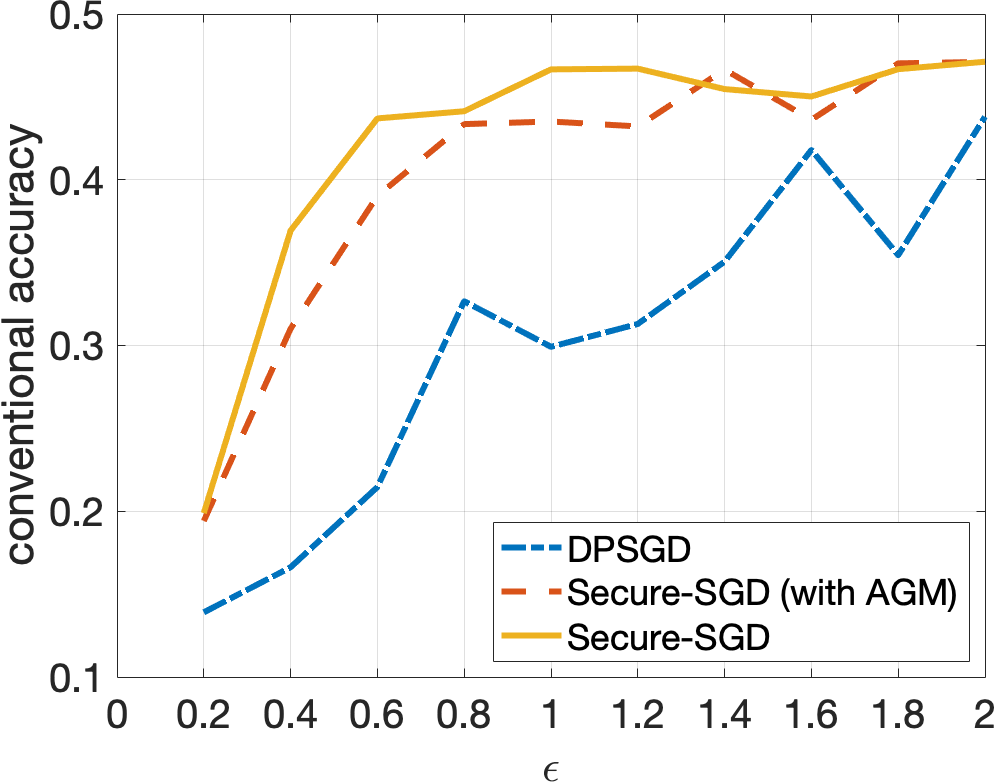} & \includegraphics[width=1.75in]{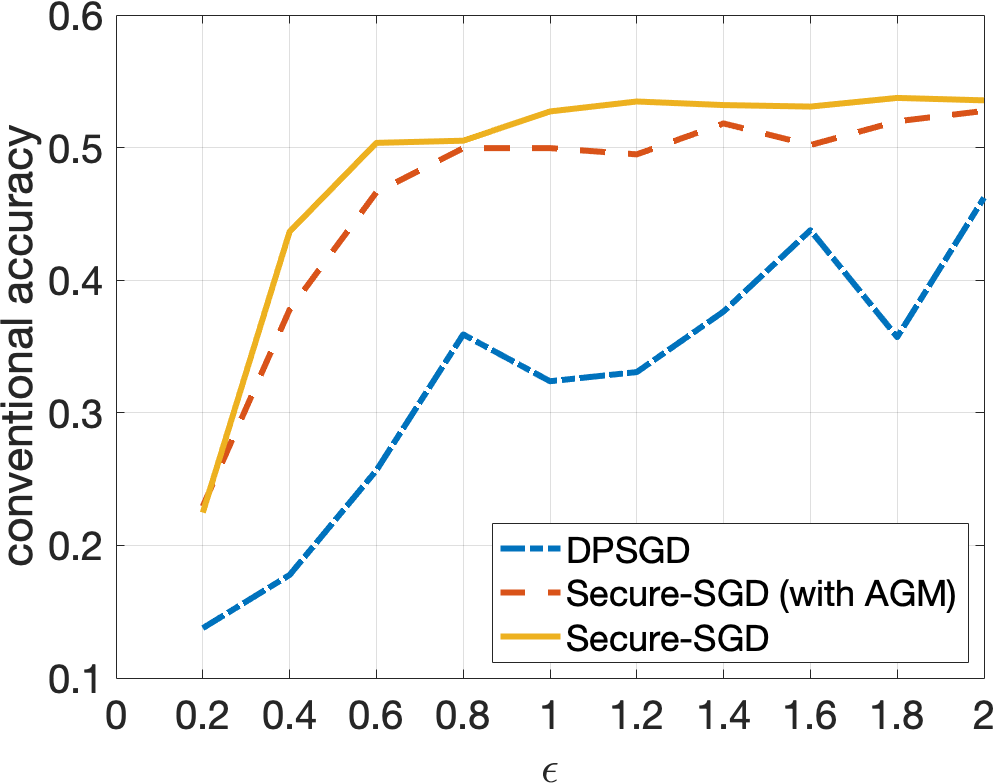} \\ [0.0cm]
\mbox{(a) I-FGSM attacks} & \mbox{(b) FGSM attacks} & \mbox{(c) MIM attacks} & \mbox{(d) MadryEtAl attacks}
\end{array}$ \vspace{-7.5pt}
\caption{Conventional accuracy on the MNIST dataset, given Secure-SGD and DPSGD, i.e., $\l_\infty(\mu_a = 0.1)$, $\epsilon_r = 4$.} \vspace{-10pt}
\label{MNIST2} 
\end{figure*}

The descent of the parameters explicitly is as: $\boldsymbol{\theta}_{t+1} \leftarrow \boldsymbol{\theta}_t - \xi_t \widetilde{g}_t$, where $\xi_t$ is a learning rate at the step $t$ (\textit{Line 16}).
The training process of our mechanism achieves both $(\epsilon, \delta)$-DP to protect the training data and provable robustness with the budgets $(\epsilon_r, \delta_r)$. In the verified testing phase (\textit{Lines 17-22}), by applying HGM and PixelDP, we derive a novel robustness bound $\mu_{max}$ for a specific input $x$ as follows: 
\begin{empheq}[box=\fbox]{align}
& \mu_{max} = \max_{\mu \in \mathbb{R}^+} \mu \text{,\ \ \ such that \ \ \ } \forall \alpha \in l_p(\mu): \nonumber
\\ 
& \hat{\mathbb{E}}_{lb} f_k(x) > e^{2\epsilon_r} \max_{i: i\neq k} \hat{\mathbb{E}}_{ub} f_{i}(x) + (1 + e^{\epsilon_r})\delta_r \nonumber 
\\
& \sigma_r = \frac{\sqrt{2}}{2\epsilon_r} (\sqrt{s} + \sqrt{s+\epsilon_r}) \Delta_f \times \mu /\epsilon_r \text{\ \ and\ \ } \epsilon_r > 0  
\label{RobustCon4} 
\end{empheq}
where $\hat{\mathbb{E}}_{lb}$ and $\hat{\mathbb{E}}_{ub}$ are the lower and upper bounds of the expected value $\hat{\mathbb{E}} f(x) = \frac{1}{N} \sum_N f(x)_N$, derived from the Monte Carlo estimation with an $\eta$-confidence, given $N$ is the number of invocations of $f(x)$ with independent draws in the noise $\gamma \leftarrow \mathcal{N}(0, \sigma_r^2 K\mathbf{r})$. Similar to \cite{Lecuyer2018}, we use Hoeffding's inequality \cite{10.2307/2282952} to bound the error in $\hat{\mathbb{E}} f(x)$. If the robustness size $\mu_{max}$ is larger than a given adversarial perturbation size $\mu_a$, the model prediction is considered consistent to that attack size. Given the relaxed budget $\epsilon_r > 0$ and the noise redistribution $K\mathbf{r}$, the search space for the robustness size $\mu_{max}$ is significantly enriched, e.g., $\epsilon_r > 1$, strengthening the robustness bound. Note that vector $\mathbf{r}$ can also be randomly drawn in the estimation of the expected value $\hat{\mathbb{E}} f(x)$.
Both fully-connected and convolution layers can be applied. Given a convolution layer, we need to ensure that the computation of each feature map is $(\epsilon_r, \delta_r)$-PixelDP, since each of them is independently computed by reading a local region of input neurons. Therefore, the sensitivity $\Delta_{f}$ can be considered the upper-bound sensitivity given any single feature map. Our algorithm is the first effort to connect DP preservation in order to protect the original training data and provable robustness in deep learning. \vspace{-5pt}


\section{Experimental Results} \vspace{-2.5pt}

We have carried out extensive experiments on two benchmark datasets, MNIST and CIFAR-10. 
Our goal is to evaluate whether our HGM significantly improves the robustness of both differentially private and non-private models under strong adversarial attacks, and whether our Secure-SGD approach retains better model utility compared with baseline mechanisms, under the same DP guarantees and protections.

\textbf{Baseline Approaches.} 
Our \textbf{HGM} and two approaches, including \textbf{HGM\_PixelDP} and \textbf{Secure-SGD}, are evaluated in comparison with state-of-the-art mechanisms in: (1) DP-preserving algorithms in deep learning, i.e., \textbf{DPSGD} \cite{Abadi}, \textbf{AdLM} \cite{NHPhanICDM17}; in (2) Provable robustness, i.e., \textbf{PixelDP} \cite{Lecuyer2018}; and (3) The Analytic Gaussian Mechanism (\textbf{AGM}) \cite{pmlr-v80-balle18a}.
To preserve DP, DPSGD injects random noise into gradients of parameters, while AdLM is a Functional Mechanism-based approach. PixelDP is one of the state-of-the-art mechanisms providing provable robustness using DP bounds. Our \textbf{HGM\_PixelDP} model simply is PixelDP with the noise bound derived from our HGM. The baseline models share the same design in our experiment. We consider the class of $l_\infty$-bounded adversaries.
Four white-box attack algorithms were used, including \textbf{FGSM}, \textbf{I-FGSM}, Momentum Iterative Method (\textbf{MIM}) \cite{DBLP:journals/corr/abs-1710-06081}, and \textbf{MadryEtAl} \cite{madry2018towards}, to draft adversarial examples $l_\infty(\mu_a)$.

\textit{MNIST:} We used two convolution layers (32 and 64 features). Each hidden neuron connects with a 5x5 unit patch. A fully-connected layer has 256 units. The batch size $m$ was set to 128, $\xi = 1.5$, $\psi = 2$, $T_\mu = 10$, and $\beta=1$.
\textit{CIFAR-10:} We used three convolution layers (128, 128, and 256 features). Each hidden neuron connects with a 3x3 unit patch in the first layer, and a 5x5 unit patch in other layers. One fully-connected layer has 256 neurons. The batch size $m$ was set to 128, $\xi = 1.5$, $\psi = 10$, $T_\mu = 3$, and $\beta=1$. Note that $\boldsymbol{\epsilon}$ is used to indicate the DP budget used to protect the training data; meanwhile, $\boldsymbol{\epsilon_r}$ is the budget for robustness.
The implementation of our mechanism is available in TensorFlow\footnote{\url{https://github.com/haiphanNJIT/SecureSGD}}. 
We apply two accuracy metrics as follows: 
\begin{small}
\begin{align}
& \textit{conventional accuracy} = \frac{\sum_{i = 1}^{|test|} isCorrect(x_i)}{|test|} \nonumber 
\\
& \textit{certified accuracy} = \frac{\sum_{i = 1}^{|test|} isCorrect(x_i) \textit{ \& } isRobust(x_i)}{|test|} \nonumber
\end{align}
\end{small}
where $|test|$ is the number of test cases, $isCorrect(\cdot)$ returns $1$ if the model makes a correct prediction (otherwise, returns 0), and $isRobust(\cdot)$ returns $1$ if the robustness size is larger than a given attack bound $\mu_a$ (otherwise, returns 0).

\textbf{HGM\_PixelDP.}
Figures \ref{Cifar1} and \ref{MNIST1} illustrate the certified accuracy under attacks of each model as a function of the adversarial perturbation $\mu_a$. Our HGM\_PixelDP notably outperforms the PixelDP model in most of the cases given the CIFAR-10 dataset. We register an  improvement of 8.63\% on average when $\epsilon_r = 8$ compared with the PixelDP, i.e., $p<8.14e-7$ (2 tail t-test). This clearly shows the effectiveness of our HGM in enhancing the robustness against adversarial examples. Regarding the MNIST data, our HGM\_PixelDP model achieves better certified accuracies when $\mu \leq 0.3$ compared with the PixelDP model. On average, our HGM\_PixelDP ($\epsilon_r = 4$) improves 4.17\% in terms of certified accuracy given $\mu_a \leq 0.3$, compared with the PixelDP, $p<5.89e-3$ (2 tail t-test). Given very strong adversarial perturbation $\mu_a > 0.3$, smaller $\epsilon_r$ usually yields better results, offering the flexibility in choosing appropriate DP budget $\epsilon_r$ for robustness given different attack magnitudes. These experimental results clearly show crucial benefits of relaxing the constraints of the privacy budget and of the heterogeneous noise distribution in our HGM. 

\textbf{Secure-SGD.} The application of our HGM in DP-preserving deep neural networks, i.e., Secure-SGD, further strengthens our observations. Figures \ref{Cifar2} and \ref{MNIST2} illustrate the certified accuracy under attacks of each model as a function of the privacy budget $\epsilon$ used to protect the training data. By incorporating HGM into DPSGD, our Secure-SGD remarkably increases the robustness of differentially private deep neural networks. In fact, our Secure-SGD with HGM outmatches DGSGP, AdLM, and the application of AGM in our Secure-SGD algorithm in most of the cases. Note that the application of AGM in our Secure-SGD does not redistribute the noise in deriving the provable robustness. In CIFAR-10 dataset, our Secure-SGD ($\epsilon_r = 8$) correspondingly acquires a 2.7\% gain ($p<1.22e-6$, 2 tail t-test), a 3.8\% gain ($p<2.16e-6$, 2 tail t-test), and a 17.75\% gain ($p<2.05e-10$, 2 tail t-test) in terms of conventional accuracy, compared with AGM in Secure-SGD, DPSGD, and AdLM algorithms. We register the same phenomenon in the MNIST dataset. On average, our Secure-GSD ($\epsilon_r = 4$) correspondingly outperforms the AGM in Secure-SGD and DPSGD with an improvement of 2.9\% ($p<8.79e-7$, 2 tail t-test) and an improvement of 10.74\% ($p<8.54e-14$, 2 tail t-test). 

\textbf{Privacy Preserving and Provable Robustness.} We also discover an original, interesting, and crucial trade-off between DP preserving to protect the training data and the provable robustness (Figures \ref{Cifar2} and \ref{MNIST2}). Given our Secure-SGD model, there is a huge improvement in terms of conventional accuracy when the privacy budget $\epsilon$ increases from 0.2 to 2 in MNIST dataset (i.e., 29.67\% on average), and from 2 to 10 in CIFAR-10 dataset (i.e., 18.17\% on average). This opens a long-term research avenue to achieve better provable robustness under strong privacy guarantees, since with strong privacy guarantees (i.e., small values of $\epsilon$), the conventional accuracies of all models are still modest. \vspace{-5pt}

\section{Conclusion} 

In this paper, we presented a Heterogeneous Gaussian Mechanism (HGM) to relax the privacy budget constraint, i.e., from $(0, 1]$ to $(0, \infty)$, and its heterogeneous noise bound. An original application of our HGM in DP-preserving mechanism with provable robustness was designed to enhance the robustness of DP deep neural networks, by introducing a novel Secure-SGD algorithm with a better robustness bound. Our model shows promising results and opens a long-term avenue to address the trade-off between DP preservation and provable robustness. In future work, we will learn how to identify and incorporate more practical Gaussian noise distributions to further improve the model accuracies under model attacks. \vspace{-20pt}

\section*{Acknowledgement} \vspace{-2.5pt}

This work is partially supported by grants DTRA HDTRA1-14-1-0055, NSF CNS-1850094, NSF CNS-1747798, NSF IIS-1502273, and NJIT Seed Grant.

\bibliographystyle{named}
\bibliography{dou,prop,thesis,paea,all,related,sigproc,bib-xintao,sigproc2,sigproc3,satcproposal}


\appendix

\section{Proof of Theorem \ref{GGaussian}}

\begin{figure}[h]
\centering
\includegraphics[width=3in]{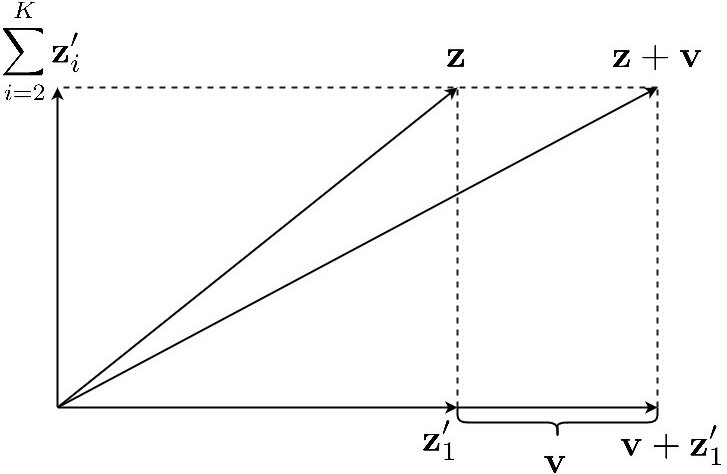}
\caption{The transformation of the hypotenuse of $\mathbf{z}+\mathbf{v}$}
\label{prooftheorem1Figure}
\end{figure}

\begin{proof} The privacy loss of the Extended Gaussian Mechanism incurred by observing an output $\mathbf{o}$ is defined as: 
\begin{equation}
\mathcal{L}( \mathbf{o}; \mathcal{M}, D, D' ) = \ln \frac{\mathrm{Pr}[\mathcal{M}\big(D, A,\sigma\big)=\mathbf{o}]}{\mathrm{Pr}[\mathcal{M}\big(D', A,\sigma\big)=\mathbf{o}]}
\end{equation}
Given $\mathbf{v} = A(D) - A(D')$, we have that 
\begin{equation}
\begin{split}
	&|\mathcal{L}(\mathbf{o}; \mathcal{M},D,D')| =\left|\ln\frac{\mathrm{Pr}[A(D)+\mathcal{N}\Big(0,\sigma^2 {\Delta}^2_A\Big)=\mathbf{o}]} {\mathrm{Pr}[A(D')+\mathcal{N}\Big(0,\sigma^2 {\Delta}^2_A \Big)=\mathbf{o}]}\right| \\
	&=\left|\ln\frac{\prod_{i=1}^K \exp\Big( -\frac{1}{2\sigma^2 {\Delta}^2_A} \big(o_i-A(D)_i\big)^2 \Big)} {\prod_{i=1}^K \exp\Big( -\frac{1}{2\sigma^2 {\Delta}^2_A} \big(o_i-A(D)_i+v_i\big)^2 \Big)}  \right|\\
	&=\frac{1}{2\sigma^2 {\Delta}^2_A}\left| \sum_{i=1}^K \big(o_i-A(D)_i\big)^2 - \big(o_i-A(D)_i+v_i\big)^2 \right| \\
	&=\frac{1}{2\sigma^2 {\Delta}^2_A} \left| \| \mathbf{z} \|^2 - \|\mathbf{z}+\mathbf{v} \|^2\right| \nonumber
\end{split}
\end{equation}
where $\mathbf{z} = \{z_i = o_i-A(D)_i\}_{i \in [1, K]}$. 

Since $\mathbf{o}-A(D)\sim \mathcal{N}\big(0,\sigma^2 {\Delta}^2_A\big)$, then $\mathbf{z} \sim \mathcal{N}\big(0,\sigma^2 {\Delta}^2_A) \big)$.
Now we will use the fact that the distribution of a spherically symmetric normal is independent of the orthogonal basis, from which its constituent normals are drawn. Then, we work in a basis that is aligned with $\mathbf{v}$. 

Let $\mathbf{b}_1,\dots,\mathbf{b}_K$ be a basis that satisfies $\|\mathbf{b}_i\|=1 ~(i\in [1, K]) $ and $\mathbf{b}_i \cdot \mathbf{b}_{i'}=0~ (i,i'\in [1, K], i \neq i')$. 
Fix such a basis $\mathbf{b}_1,\dots,\mathbf{b}_K$, we draw $\mathbf{z}$ by first drawing signed lengths $\lambda_i \sim \mathcal{N}\big(0,\sigma^2 {\Delta}^2_A \big) ~( i \in [1, K])$. 
Then, let $\mathbf{z}'_i = \lambda_i \mathbf{b}_i$ and $\mathbf{z}=\sum_{i=1}^K \mathbf{z}'_i$.  Without loss of generality, let us assume that $\mathbf{b}_1$ is parallel to $\mathbf{v}$. 
Consider that the triangle with base $\mathbf{v}+\mathbf{z}'_1$ and the edge $\sum_{i=2}^K \mathbf{z}'_i$ is orthogonal to $\mathbf{v}$. The hypotenuse of this triangle is $\mathbf{z}+\mathbf{v}$ (Figure \ref{prooftheorem1Figure}). Then we have
\begin{equation}
\|\mathbf{z}+\mathbf{v} \|^2 = \| \mathbf{v}+\mathbf{z}'_1 \|^2+ \sum_{i=2}^K \|\mathbf{z}'_i\|^2, \text{\ \ \ }
\|\mathbf{z}\|^2= \sum_{i=1}^K \|\mathbf{z}'_i\|^2 \nonumber
\end{equation}
Since $\mathbf{v}$ is parallel to $\mathbf{z}'_1$, we have $\|\mathbf{v} + \mathbf{z}'_1 \|^2=(\|\mathbf{v}\|+\lambda_1)^2$. Then we have
\begin{equation}
\begin{split}
&|\mathcal{L}(\mathbf{o}; \mathcal{M},D,D')| = \frac{1}{2\sigma^2 {\Delta}^2_A} \left| \| \mathbf{z} \|^2 - \|\mathbf{z}+\mathbf{v} \|^2\right|\\
&=\frac{1}{2\sigma^2 {\Delta}^2_A} \left| \sum_{i=1}^K \|\mathbf{z}'_i\|^2 -(\|\mathbf{z}'_1+\mathbf{v}\|^2+\sum_{i=2}^K \|\mathbf{z}'_i\|^2) \right|\\
&=\frac{1}{2\sigma^2 {\Delta}^2_A} \left| \|\mathbf{z}'_1+\mathbf{v}\|^2- \|\mathbf{z}'_1\|^2 \right|\\
&=\frac{1}{2\sigma^2 {\Delta}^2_A}  \left|  (\|\mathbf{v}\|+\lambda_1)^2-\lambda_1^2 \right| 
=\frac{1}{2\sigma^2 {\Delta}^2_A}  \left|  \|\mathbf{v}\|^2+2\lambda_1 \|\mathbf{v}\| \right| \\
&\le \frac{1}{2\sigma^2 {\Delta}^2_A}  \left| {\Delta}^2_A  + 2|\lambda_1| {\Delta}_A \right|
=\frac{1}{2\sigma^2} \left| 1 + \frac{2|\lambda_1|}{{\Delta}_A}\right|  \nonumber
\end{split}
\end{equation}
By bounding the privacy loss by $\epsilon ~ (\epsilon>0)$, we have 
\begin{equation}
\begin{split}
&|\mathcal{L}(\mathbf{o}; \mathcal{M},D,D')| \le \frac{1}{2\sigma^2} \left| 1 + \frac{2|\lambda_1|}{{\Delta}_A}\right| \le \epsilon\\
\Leftrightarrow &-2\sigma^2 \epsilon \le 1 + \frac{2|\lambda_1|}{{\Delta}_A} \le 2\sigma^2 \epsilon \\
\Leftrightarrow & |\lambda_1| \le  \frac{{\Delta}_A}{2} (2\sigma^2\epsilon-1) \nonumber
\end{split}
\end{equation}
Let $\lambda_{max}=\frac{{\Delta}_A}{2} (2\sigma^2\epsilon-1)$. To ensure the privacy loss is bounded by $\epsilon$ with probability at least $1-\delta$, we require
\begin{equation}
\begin{split}
\mathrm{Pr} (|\lambda_1| \le \lambda_{max}) \ge 1-\delta
\end{split}
\end{equation}
Recall that $\lambda_1\sim \mathcal{N}(0,\sigma^2 {\Delta}^2_A)$,
we have that
\begin{equation}
\mathrm{Pr}  (|\lambda_1| \le \lambda_{max}) = 1-2\mathrm{Pr} (\lambda_1 > \lambda_{max})
\end{equation}
Then, we have
\begin{equation}
\begin{split}
&\mathrm{Pr} (|\lambda_1| \le \lambda_{max}) \ge 1-\delta\\
\Leftrightarrow & 1-2\mathrm{Pr} (\lambda_1 > \lambda_{max}) \ge 1-\delta
\Leftrightarrow \mathrm{Pr} (\lambda_1 > \lambda_{max}) \le \frac{\delta}{2} \nonumber
\end{split}
\end{equation}
Next we will use the tail bound:
$\mathrm{Pr}(\lambda_1>t)\le \frac{\sigma {\Delta}_A}{\sqrt{2\pi}}e^{-\frac{t^2}{2\sigma^2 {\Delta}^2_A}}$.
We require:
\begin{equation}
\begin{split}
&\frac{\sigma {\Delta}_A}{\sqrt{2\pi}t}e^{-\frac{t^2}{2\sigma^2 {\Delta}^2_A}} \le \frac{\delta}{2} 
\Leftrightarrow \frac{\sigma {\Delta}_A}{t}e^{-\frac{t^2}{2\sigma^2 {\Delta}^2_A}} \le \frac{\sqrt{2\pi}\delta}{2}\\
\Leftrightarrow & \frac{t}{\sigma {\Delta}_A} e^{\frac{t^2}{2\sigma^2 {\Delta}^2_A}} \ge \sqrt{\frac{2}{\pi}}\frac{1}{\delta}\\
\end{split}
\end{equation}

Taking $t=\lambda_{max}=\frac{{\Delta}_A}{2} (2\sigma^2\epsilon-1)$, we have that
\begin{equation}
\begin{split}
& \frac{t}{\sigma {\Delta}_A} e^{\frac{t^2}{2\sigma^2 {\Delta}^2_A}} \ge \sqrt{\frac{2}{\pi}}\frac{1}{\delta}
\Leftrightarrow \frac{2\sigma^2\epsilon-1}{2\sigma} e^{\frac{1}{2}\big(\frac{2\sigma^2\epsilon-1}{2\sigma}\big)^2} \ge \sqrt{\frac{2}{\pi}}\frac{1}{\delta}\\
& \Leftrightarrow \ln \frac{2\sigma^2\epsilon-1}{2\sigma} + \frac{1}{2} \big(\frac{2\sigma^2\epsilon-1}{2\sigma}\big)^2 \ge \ln ( \sqrt{\frac{2}{\pi}}\frac{1}{\delta}) \nonumber
\end{split}
\end{equation}

We will ensure the above inequality by requiring: \textbf{(1)} $ \ln \frac{2\sigma^2\epsilon-1}{2\sigma} \ge 0$
, and \textbf{(2)} $\frac{1}{2} \big(\frac{2\sigma^2\epsilon-1}{2\sigma}\big)^2 \ge \ln ( \sqrt{\frac{2}{\pi}}\frac{1}{\delta})$. 
\begin{equation}
\begin{split}
&\ln \frac{2\sigma^2\epsilon-1}{2\sigma} \ge 0 \Leftrightarrow \frac{2\sigma^2\epsilon-1}{2\sigma} \ge 1\\
& \Leftrightarrow 2\sigma^2\epsilon - 2\sigma -1 \ge 0
\end{split}
\label{Eq14}
\end{equation}
We can ensure this inequality (Eq. \ref{Eq14}) by setting:
\begin{equation}
\begin{split}
\sigma \ge \frac{1+\sqrt{1+2\epsilon}}{2\epsilon} ~\text{(\textbf{Condition 1})}.
\end{split}
\end{equation}

Let $s=\ln(\sqrt{\frac{2}{\pi}}\frac{1}{\delta})$. If $s<0$, the second requirement will always be satisfied, and we only need to choose $\sigma$ satisfying the \textbf{Condition 1}. When $s \ge 0$, since we already ensure $\frac{2\sigma^2\epsilon-1}{2\sigma}\ge 1$, we have that 
\begin{equation}
\begin{split}
&\frac{1}{2} \big(\frac{2\sigma^2\epsilon-1}{2\sigma}\big)^2 \ge s
\Leftrightarrow \frac{2\sigma^2\epsilon-1}{2\sigma} \ge \sqrt{2s}\\
& \Leftrightarrow 2\sigma^2\epsilon - 2\sigma \sqrt{2s} -1 \ge 0
\end{split}
\end{equation}
We can ensure the above inequality by choosing:
\begin{equation}
\begin{split}
		\sigma \ge \frac{\sqrt{2}}{2\epsilon} (\sqrt{s} + \sqrt{s+\epsilon}) ~\text{(\textbf{Condition 2})}
\end{split}
\end{equation}
Based on the proof above, now we know that to ensure the privacy loss $|\mathcal{L}(\mathbf{o}; \mathcal{M},D,D')|$ bounded by $\epsilon$ with probability at least $1-\delta$, we require:
\begin{equation}
\begin{split}
&\sigma \ge \frac{1+\sqrt{1+2\epsilon}}{2\epsilon} ~\text{(\textbf{Condition 1})};\\
&\sigma \ge \frac{\sqrt{2}}{2\epsilon} (\sqrt{s} + \sqrt{s+\epsilon}) ~\text{(\textbf{Condition 2})}
\end{split}
\end{equation} 
To compare \textbf{Condition 2} and \textbf{Condition 1}, we have that
\begin{equation}
\begin{split}
&  \frac{\sqrt{2}}{2\epsilon} (\sqrt{s} + \sqrt{s+\epsilon})  >  \frac{1+\sqrt{1+2\epsilon}}{2\epsilon} \\
& \Leftrightarrow \sqrt{2s} + \sqrt{2s+2\epsilon} > 1 + \sqrt{1+2\epsilon}\\
& \Leftrightarrow \sqrt{2s} > 1\Leftrightarrow s > \frac{1}{2} 
\Leftrightarrow \ln( \sqrt{ \frac{2}{\pi}} \frac{1}{\delta}) >  \frac{1}{2}\\
& \Leftrightarrow \delta < \sqrt{ \frac{2}{\pi}}  e^{-\frac{1}{2}} \approx 0.48.
\end{split}
\end{equation}
Since $\delta$ usually is a very small number, i.e., $(1e$-$5 \ll 0.48)$, without loss of generality, we can assume that \textbf{Condition 2} always implies \textbf{Condition 1} in practice. To ensure the privacy loss bounded by $\epsilon$ with probability at least $1-\delta$, only \textbf{Condition 2} needs to be satisfied: 
\begin{align}
& \mathrm{Pr}[\mathcal{M}\big(D, A,\sigma\big)=\mathbf{o}] \leq e^\epsilon \mathrm{Pr}[\mathcal{M}\big(D', A,\sigma\big)=\mathbf{o}] + \delta \nonumber 
\\
&\text{\ \ with \ \ } \sigma \ge \frac{\sqrt{2}}{2\epsilon} (\sqrt{s} + \sqrt{s+\epsilon}), \text{\ \ and \ \ } \epsilon > 0
\end{align}
In this proof, the noise $\mathcal{N}(0, \sigma^2\Delta_A^2)$ is injected into the model. If we set $\sigma \ge \frac{\sqrt{2}\Delta_A}{2\epsilon} (\sqrt{s} + \sqrt{s+\epsilon})$, then the noise becomes $\mathcal{N}(0, \sigma^2)$. Consequently, Theorem \ref{GGaussian} does hold.
\end{proof}

\section{Proof of Theorem \ref{HGaussian}}

\begin{proof} The privacy loss of the Heterogeneous Gaussian Mechanism incurred by observing an output $\mathbf{o}$ is defined as: 
\begin{equation}
\mathcal{L}( \mathbf{o}; \mathcal{M}, D, D' ) = \ln \frac{\mathrm{Pr}[\mathcal{M}\big(D, A, \mathbf{r}, \sigma\big)=\mathbf{o}]}{\mathrm{Pr}[\mathcal{M}\big(D', A,\mathbf{r},\sigma\big)=\mathbf{o}]}
\end{equation}
Given $\mathbf{v} = A(D) - A(D')$, we have that 
\begin{equation}
\begin{split}
	&|\mathcal{L}(\mathbf{o}; \mathcal{M},D,D')|\\
	&=\left|\ln\frac{\mathrm{Pr}[\mathcal{M}\big(D,A,\mathbf{r},\sigma\big)=\mathbf{o}]}{\mathrm{Pr}[\mathcal{M}\big(D',A,\mathbf{r},\sigma\big)=\mathbf{o}]}\right|\\
	&=\left|\ln\frac{\mathrm{Pr}[A(D)+\mathcal{N}\Big(0,\sigma^2 \Delta^2_A K \mathbf{r}\Big)=\mathbf{o}]} {\mathrm{Pr}[A(D')+\mathcal{N}\Big(0,\sigma^2 \Delta^2_A K \mathbf{r}\Big)=\mathbf{o}]}\right| \\
	&=\left|\ln\frac{\prod_{k=1}^K \exp\Big( -\frac{1}{2\sigma^2 {\Delta}^2_A Kr_k} \big(o_k-A(D)_k\big)^2 \Big)} {\prod_{i=1}^K \exp\Big( -\frac{1}{2\sigma^2 {\Delta}^2_A Kr_k} \big(o_k-A(D)_k+v_k\big)^2 \Big)}  \right|\\
	&=\frac{1}{2\sigma^2 \Delta^2_A}\left| \sum_{k=1}^K \frac{\big(o_k-A(D)_k\big)^2 - \big(o_k-A(D)_k+v_k\big)^2}{Kr_k} \right| \nonumber
\end{split}
\end{equation}
Let $\mathbf{z}$ be a $K$-dimensional vector that satisfies $z_k= \frac{o_k-A(D)_k}{\sqrt{Kr_k}} ~(k \in [K]).$
Let $\mathbf{v}'$ be a $K$-dimensional vector that satisfies $v'_k=\frac{v_k}{\sqrt{Kr_k}}  ~(k \in [K])$. Then we have that 
\begin{equation}
\begin{split}
&|\mathcal{L}(\mathbf{o}; \mathcal{M},D,D')| \\
&=\frac{1}{2\sigma^2 \Delta^2_A}\left| \sum_{k=1}^K \frac{\big(o_k-A(D)_k\big)^2 - \big(o_k-A(D)_k+v_k\big)^2}{Kr_k} \right| \\
&=\frac{1}{2\sigma^2 \Delta^2_A}\left| \sum_{k=1}^K \big( z_k^2-(z_k+v'_k)^2 \big)\right|\\
&= \frac{1}{2\sigma^2 \Delta^2_A} \left| \| \mathbf{z} \|^2 - \|\mathbf{z}+\mathbf{v}' \|^2\right| \nonumber
\end{split}
\end{equation}

Since $\mathbf{o}-A(D)\sim \mathcal{N}\big(0,\sigma^2 {\Delta}^2_A K\mathbf{r}\big)$, then $\mathbf{z} \sim \mathcal{N}\big(0,\sigma^2 \Delta^2_A \big)$.
Now we will use the fact that the distribution of a spherically symmetric normal is independent of the orthogonal basis from which its constituent normals are drawn. Then, we work in a basis that is aligned with $\mathbf{v}'$.

Let $\mathbf{b}_1,\dots,\mathbf{b}_K$ be a basis that satisfies $\|\mathbf{b}_k\|=1 ~(k\in [K]) $ and $\mathbf{b}_k \cdot \mathbf{b}_{k'}=0~ (k,k'\in [K], k\neq k')$. 
Fix such a basis $\mathbf{b}_1,\dots,\mathbf{b}_K$, we draw $\mathbf{z}$ by first drawing signed lengths $\lambda_k \sim \mathcal{N}\big(0,\sigma^2 {\Delta}^2_A \big) ~( k\in [K])$.
Then, let $\mathbf{z}'_k=\lambda_k\mathbf{b}_k$, and finally let $\mathbf{z}=\sum_{k=1}^K \mathbf{z}'_i$. Assume without loss of generality  that $\mathbf{b}_1$ is parallel to $\mathbf{v}'$. 
Consider that the right triangle with base $\mathbf{v}+\mathbf{z}'_1$ and edge $\sum_{k=2}^K \mathbf{z}'_i$ orthogonal to $\mathbf{v}$. The hypotenuse of this triangle is $\mathbf{z}+\mathbf{v}$ (Figure \ref{prooftheorem1Figure}). Then we have 
\begin{equation}
\|\mathbf{z}+\mathbf{v} \|^2 = \| \mathbf{v}+\mathbf{z}'_1 \|^2+ \sum_{k=2}^K \|\mathbf{z}'_k\|^2, \text{\ \ \ }
\|\mathbf{z}\|^2= \sum_{k=1}^M \|\mathbf{z}'_k\|^2 \nonumber
\end{equation}
Since $\mathbf{v}$ is parallel to $\mathbf{z}'_1$, we have $\|\mathbf{z}'_1+\mathbf{v} \|^2=(\|\mathbf{v}\|+\lambda_1)^2$. Then we have
\begin{equation}
\begin{split}
&|\mathcal{L}(\mathbf{o}; \mathcal{M},D,D')| = \frac{1}{2\sigma^2 {\Delta}^2_A} \left| \| \mathbf{z} \|^2 - \|\mathbf{z}+\mathbf{v} \|^2\right|\\
&=\frac{1}{2\sigma^2 {\Delta}^2_A} \left| \sum_{i=1}^K \|\mathbf{z}'_i\|^2 -(\|\mathbf{z}'_1+\mathbf{v}\|^2+\sum_{i=2}^K \|\mathbf{z}'_i\|^2) \right|\\
&=\frac{1}{2\sigma^2 {\Delta}^2_A} \left| \|\mathbf{z}'_1+\mathbf{v}\|^2- \|\mathbf{z}'_1\|^2 \right|\\
&=\frac{1}{2\sigma^2 {\Delta}^2_A}  \left|  (\|\mathbf{v}\|+\lambda_1)^2-\lambda_1^2 \right| 
=\frac{1}{2\sigma^2 {\Delta}^2_A}  \left|  \|\mathbf{v}\|^2+2\lambda_1 \|\mathbf{v}\| \right| \\
&\le \frac{1}{2\sigma^2 {\Delta}^2_A}  \left| {\Delta}^2_A  + 2|\lambda_1| {\Delta}_A \right|
=\frac{1}{2\sigma^2} \left| 1 + \frac{2|\lambda_1|}{{\Delta}_A}\right|  \nonumber
\end{split}
\end{equation}
By bounding the privacy loss by $\epsilon ~ (\epsilon>0)$, we have 
\begin{equation}
\begin{split}
&|\mathcal{L}(\mathbf{o}; \mathcal{M},D,D')| \le \frac{1}{2\sigma^2} \left| 1 + \frac{2|\lambda_1|}{{\Delta}_A}\right| \le \epsilon\\
\Leftrightarrow &-2\sigma^2 \epsilon \le 1 + \frac{2|\lambda_1|}{{\Delta}_A} \le 2\sigma^2 \epsilon \\
\Leftrightarrow & |\lambda_1| \le  \frac{{\Delta}_A}{2} (2\sigma^2\epsilon-1) \nonumber
\end{split}
\end{equation}
Let $\lambda_{max}=\frac{{\Delta}_A}{2} (2\sigma^2\epsilon-1)$. To ensure the privacy loss is bounded by $\epsilon$ with probability at least $1-\delta$, we require
\begin{equation}
\begin{split}
\mathrm{Pr} (|\lambda_1| \le \lambda_{max}) \ge 1-\delta
\end{split}
\end{equation}
Recall that $\lambda_1\sim \mathcal{N}(0,\sigma^2 {\Delta}^2_A)$,
we have that
\begin{equation}
\mathrm{Pr}  (|\lambda_1| \le \lambda_{max}) = 1-2\mathrm{Pr} (\lambda_1 > \lambda_{max})
\end{equation}
Then, we have
\begin{equation}
\begin{split}
&\mathrm{Pr} (|\lambda_1| \le \lambda_{max}) \ge 1-\delta\\
\Leftrightarrow & 1-2\mathrm{Pr} (\lambda_1 > \lambda_{max}) \ge 1-\delta
\Leftrightarrow \mathrm{Pr} (\lambda_1 > \lambda_{max}) \le \frac{\delta}{2} \nonumber
\end{split}
\end{equation}
Next we will use the tail bound:
$\mathrm{Pr}(\lambda_1>t)\le \frac{\sigma {\Delta}_A}{\sqrt{2\pi}}e^{-\frac{t^2}{2\sigma^2 {\Delta}^2_A}}$.
We require:
\begin{equation}
\begin{split}
&\frac{\sigma {\Delta}_A}{\sqrt{2\pi}t}e^{-\frac{t^2}{2\sigma^2 {\Delta}^2_A}} \le \frac{\delta}{2} 
\Leftrightarrow \frac{\sigma {\Delta}_A}{t}e^{-\frac{t^2}{2\sigma^2 {\Delta}^2_A}} \le \frac{\sqrt{2\pi}\delta}{2}\\
\Leftrightarrow & \frac{t}{\sigma {\Delta}_A} e^{\frac{t^2}{2\sigma^2 {\Delta}^2_A}} \ge \sqrt{\frac{2}{\pi}}\frac{1}{\delta}\\
\end{split}
\end{equation}

Taking $t=\lambda_{max}=\frac{{\Delta}_A}{2} (2\sigma^2\epsilon-1)$, we have that
\begin{equation}
\begin{split}
& \frac{t}{\sigma {\Delta}_A} e^{\frac{t^2}{2\sigma^2 {\Delta}^2_A}} \ge \sqrt{\frac{2}{\pi}}\frac{1}{\delta}
\Leftrightarrow \frac{2\sigma^2\epsilon-1}{2\sigma} e^{\frac{1}{2}\big(\frac{2\sigma^2\epsilon-1}{2\sigma}\big)^2} \ge \sqrt{\frac{2}{\pi}}\frac{1}{\delta}\\
& \Leftrightarrow \ln \frac{2\sigma^2\epsilon-1}{2\sigma} + \frac{1}{2} \big(\frac{2\sigma^2\epsilon-1}{2\sigma}\big)^2 \ge \ln ( \sqrt{\frac{2}{\pi}}\frac{1}{\delta}) \nonumber
\end{split}
\end{equation}

We will ensure the above inequality by requiring: \textbf{(1)} $ \ln \frac{2\sigma^2\epsilon-1}{2\sigma} \ge 0$
, and \textbf{(2)} $\frac{1}{2} \big(\frac{2\sigma^2\epsilon-1}{2\sigma}\big)^2 \ge \ln ( \sqrt{\frac{2}{\pi}}\frac{1}{\delta})$. 
\begin{equation}
\begin{split}
&\ln \frac{2\sigma^2\epsilon-1}{2\sigma} \ge 0 \Leftrightarrow \frac{2\sigma^2\epsilon-1}{2\sigma} \ge 1\\
& \Leftrightarrow 2\sigma^2\epsilon - 2\sigma -1 \ge 0
\end{split}
\label{Eq25}
\end{equation}
We can ensure this inequality (Eq. \ref{Eq25}) by setting:
\begin{equation}
\begin{split}
\sigma \ge \frac{1+\sqrt{1+2\epsilon}}{2\epsilon} ~\text{(\textbf{Condition 1})}.
\end{split}
\end{equation}

Let $s=\ln(\sqrt{\frac{2}{\pi}}\frac{1}{\delta})$. If $s<0$, the second requirement will always be satisfied, and we only need to choose $\sigma$ satisfying the \textbf{Condition 1}. When $s \ge 0$, since we already ensure $\frac{2\sigma^2\epsilon-1}{2\sigma}\ge 1$, we have that 
\begin{equation}
\begin{split}
&\frac{1}{2} \big(\frac{2\sigma^2\epsilon-1}{2\sigma}\big)^2 \ge s
\Leftrightarrow \frac{2\sigma^2\epsilon-1}{2\sigma} \ge \sqrt{2s}\\
& \Leftrightarrow 2\sigma^2\epsilon - 2\sigma \sqrt{2s} -1 \ge 0
\end{split}
\end{equation}
We can ensure the above inequality by choosing:
\begin{equation}
\begin{split}
		\sigma \ge \frac{\sqrt{2}}{2\epsilon} (\sqrt{s} + \sqrt{s+\epsilon}) ~\text{(\textbf{Condition 2})}
\end{split}
\end{equation}
Based on the proof above, now we know that to ensure the privacy loss $|\mathcal{L}(\mathbf{o}; \mathcal{M},D,D')|$ bounded by $\epsilon$ with probability at least $1-\delta$, we require:
\begin{equation}
\begin{split}
&\sigma \ge \frac{1+\sqrt{1+2\epsilon}}{2\epsilon} ~\text{(\textbf{Condition 1})};\\
&\sigma \ge \frac{\sqrt{2}}{2\epsilon} (\sqrt{s} + \sqrt{s+\epsilon}) ~\text{(\textbf{Condition 2})}
\end{split}
\end{equation} 
To compare \textbf{Condition 2} and \textbf{Condition 1}, we have that
\begin{equation}
\begin{split}
&  \frac{\sqrt{2}}{2\epsilon} (\sqrt{s} + \sqrt{s+\epsilon})  >  \frac{1+\sqrt{1+2\epsilon}}{2\epsilon} \\
& \Leftrightarrow \sqrt{2s} + \sqrt{2s+2\epsilon} > 1 + \sqrt{1+2\epsilon}\\
& \Leftrightarrow \sqrt{2s} > 1\Leftrightarrow s > \frac{1}{2} 
\Leftrightarrow \ln( \sqrt{ \frac{2}{\pi}} \frac{1}{\delta}) >  \frac{1}{2}\\
& \Leftrightarrow \delta < \sqrt{ \frac{2}{\pi}}  e^{-\frac{1}{2}} \approx 0.48.
\end{split}
\end{equation}
Since $\delta$ usually is a very small number, i.e., $(1e$-$5 \ll 0.48)$, without loss of generality, we can assume that \textbf{Condition 2} always implies \textbf{Condition 1} in practice. To ensure the privacy loss bounded by $\epsilon$ with probability at least $1-\delta$, only \textbf{Condition 2} needs to be satisfied: 
\begin{align}
& \mathrm{Pr}[\mathcal{M}\big(D, A,\mathbf{r}, \sigma\big)=\mathbf{o}] \leq e^\epsilon \mathrm{Pr}[\mathcal{M}\big(D', A,\mathbf{r}, \sigma\big)=\mathbf{o}] + \delta \nonumber 
\\
&\text{\ \ with \ \ } \sigma \ge \frac{\sqrt{2}}{2\epsilon} (\sqrt{s} + \sqrt{s+\epsilon}), \text{\ \ and \ \ } \epsilon > 0
\end{align}
In this proof, the noise $\mathcal{N}(0, \sigma^2\Delta_A^2K\mathbf{r})$ is injected into the model. If we set $\sigma \ge \frac{\sqrt{2}\Delta_A}{2\epsilon} (\sqrt{s} + \sqrt{s+\epsilon})$, then the noise becomes $\mathcal{N}(0, \sigma^2K\mathbf{r})$. Consequently, Theorem \ref{HGaussian} does hold.

\end{proof}

\end{document}